%% file: main.tex
\documentclass[lettersize,journal]{IEEEtran}
\usepackage{amsmath,amsfonts}
\usepackage{algorithmic}
\usepackage{algorithm}
\usepackage{array}
\usepackage[caption=false,font=normalsize,labelfont=sf,textfont=sf]{subfig}
\usepackage{textcomp}
\usepackage{stfloats}
\usepackage{url}
\usepackage{verbatim}
\usepackage{graphicx}
\usepackage{cite}
\usepackage{booktabs}
\usepackage{xcolor}
\usepackage{tikz}
\usepackage{hyperref}
\usepackage{multirow}
\usepackage{tabularx}

\newcommand{\techName}[1]{\textit{ManiScope}}
\newcommand{\HolderView}[1]{\textit{Holding Distribution View}}
\newcommand{\BehaviorView}[1]{\textit{Behavior Detail View}}
\newcommand{\PriceView}[1]{\textit{Manipulation View}}

\definecolor{CaptionColor}{RGB}{89, 91, 97}
\newcommand*\captionID[1]{\tikz[baseline=(char.base)]{
            \node[shape=rectangle,fill=white,text=CaptionColor,draw=CaptionColor,
      line width=1pt,inner sep=1.2pt,minimum size=8pt,rounded corners=1pt] (char) {\textbf{#1}}}}

\begin{document}

\title{\techName{}: LLM-Assisted Visual Analytics of Cryptocurrency Manipulation Risk}

\author{
Xiaolin Wen, 
Feng Liang, 
Yuanye Ma,
Qishuang Fu,
Zhengyu Sun,
Feng Zhu,
Can Liu, 
and Yong Wang
\thanks{X. Wen, Y. Ma, Z. Sun, F. Liang, C. Liu, and Y. Wang are with Nanyang Technological University. Email: {xiaolin004, ma0002ye, zhengyu003}@e.ntu.edu.sg, 
{feng.liang, can.liu,  yong-wang}@ntu.edu.sg. }
\thanks{Q. Fu is with Monash University. Email: qishuang.fu@monash.edu.}
\thanks{F. Zhu is with GoPlus Security. Email:  zhupiter777@gmail.com.}
}

\maketitle

\begin{abstract}
Cryptocurrency markets are vulnerable to trade-based manipulation, such as wash trading, which can distort price signals and mislead investors. Prior research has mainly focused on detecting manipulation using fixed rules or labeled examples, offering limited flexibility and interpretability for assessing potential risks. Existing visual analytics tools can reveal basic manipulation-related signals, such as token distribution, but still require substantial manual effort to integrate holder relationships, suspicious behaviors, and market dynamics for risk assessment.
To address these limitations, we propose \techName{}, an LLM-assisted visual analytics system for analyzing trade-based manipulation risks in cryptocurrency markets. \techName{} provides coordinated views of token distributions, holder relationships, detailed holder behaviors, price dynamics, and suspicious trading patterns. 
To further enhance user analysis, \techName{} introduces a human--LLM collaborative visual analytics framework.
Rather than acting as a basic reactive LLM assistant, the framework positions the LLM as a co-analyst that
infers users' analytical intent and emerging hypotheses from interaction context and surfaces relevant visual, statistical, and synthesized evidence for hypothesis evaluation. 
This design reduces repetitive inspection and strengthens evidence-based reasoning.
We evaluate \techName{} through two case studies and a user study with 12 experienced cryptocurrency practitioners. The results suggest that \techName{} supports effective risk assessment of manipulation, reduces manual effort in evidence-seeking, and organizes findings around user hypotheses.
\end{abstract}

\begin{IEEEkeywords}
Cryptocurrency manipulation, visual analytics, human-LLM collaboration.
\end{IEEEkeywords}

\input{source/1_intro}
\input{source/2_related_work}
\input{source/3_before_method}
\input{source/4_method}
\input{source/5_evaluation}
\input{source/6_dicussion_conclusion}

\bibliographystyle{IEEEtran}
\bibliography{main}

\vspace{-33pt}
\begin{IEEEbiography}[{\includegraphics[width=1in,height=1.25in,clip,keepaspectratio]{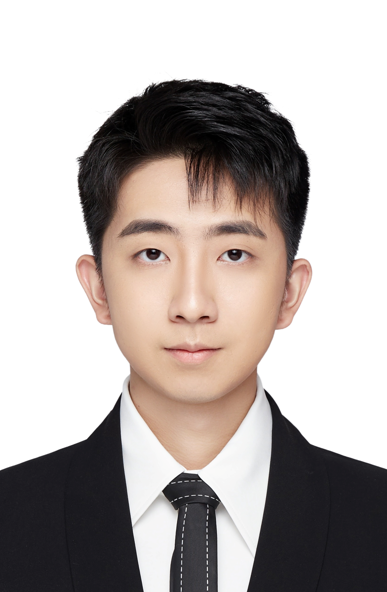}}]{Xiaolin Wen}
is currently a Ph.D student in the College of Computing and Data Science, Nanyang Technological University (NTU). 
His research interests mainly focus on visualization for FinTech and LLM-assisted design study.
He received his master's degree in Computer Science and Technology from Sichuan University in 2023 and his dual bachelor's degree in Computer Science and Financial Engineering from Sichuan University in 2016.
For more information, kindly visit \url{https://wenxiaolin.com/}.
\end{IEEEbiography}

\vspace{-33pt}

\begin{IEEEbiography}[{\includegraphics[width=1in,height=1.25in,clip,keepaspectratio]{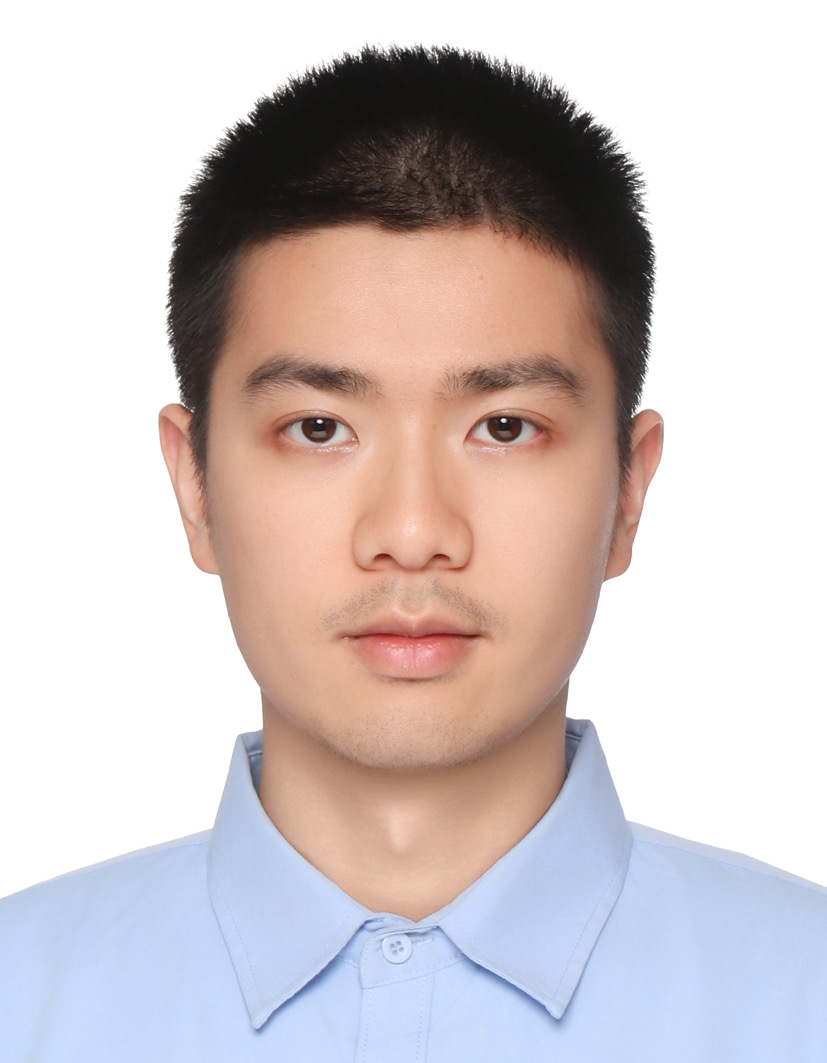}}]{Feng Liang} is currently a research engineer in the College of Computing and Data Science, Nanyang Technological University. His research interests mainly focus on Visualization for Quantum Computing and AI for Science. He received his master's degree from KAUST in 2022 and his bachelor's degree from SUSTech in 2017. For more information, kindly visit \url{https://fengliang.io/}.
\end{IEEEbiography}

\vspace{-33pt}

\begin{IEEEbiography}[{\includegraphics[width=1in,height=1.25in,clip,keepaspectratio]{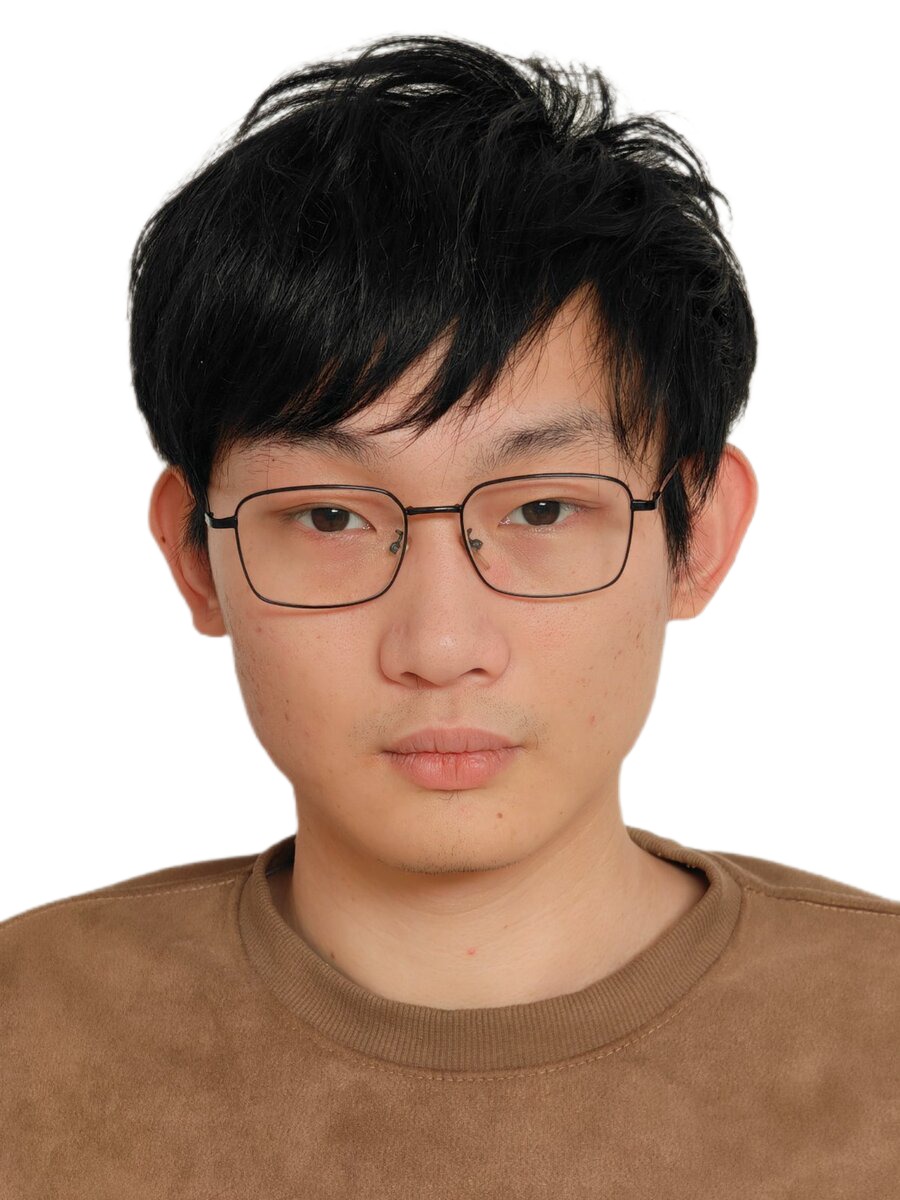}}]{Yuanye Ma} is currently a master's student in Blockchain at Nanyang Technological University. His research interests include blockchain security, cryptocurrency anti-money laundering, and on-chain transaction analysis. He received his bachelor's degree in Computer Science from the University of Bristol in 2024. For more information, kindly visit https://github.com/MerlinGert.
\end{IEEEbiography}

\vspace{-33pt}

\begin{IEEEbiography}[{\includegraphics[width=1in,height=1.25in,clip,keepaspectratio]{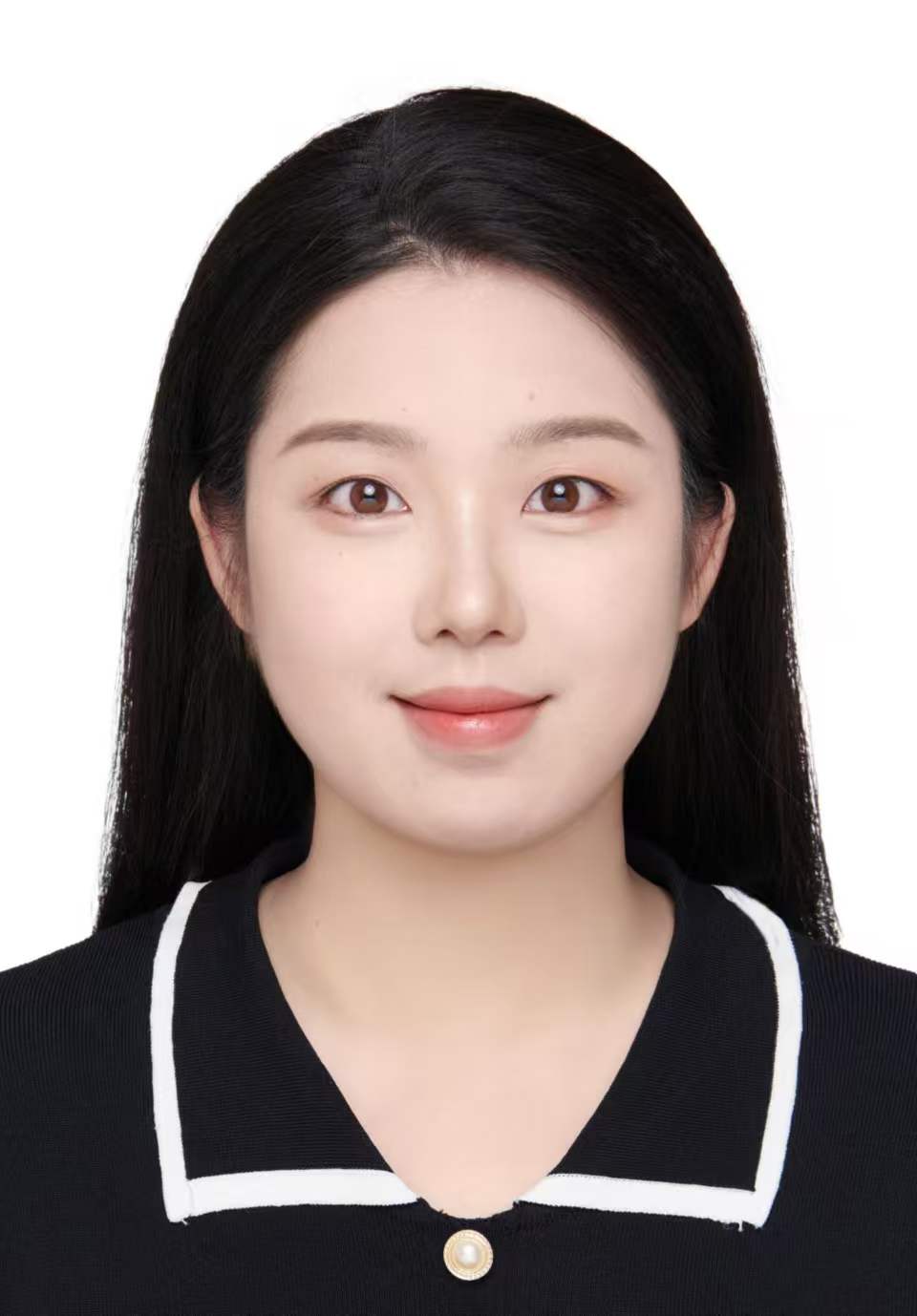}}]{Qishuang Fu} is currently a Ph.D. student in the Faculty of Information Technology, Monash University. Her research interests mainly focus on blockchain security and cryptocurrency anti-money laundering. She received her master's degree in Software Engineering from Sun Yat-sen University in 2024 and her bachelor's degree in Automation from Dalian University of Technology in 2021. For more information, kindly visit \url{https://fuqishuang228.github.io/}.
\end{IEEEbiography}

\vspace{-33pt}

\begin{IEEEbiography}[{\includegraphics[width=1in,height=1.25in,clip,keepaspectratio]{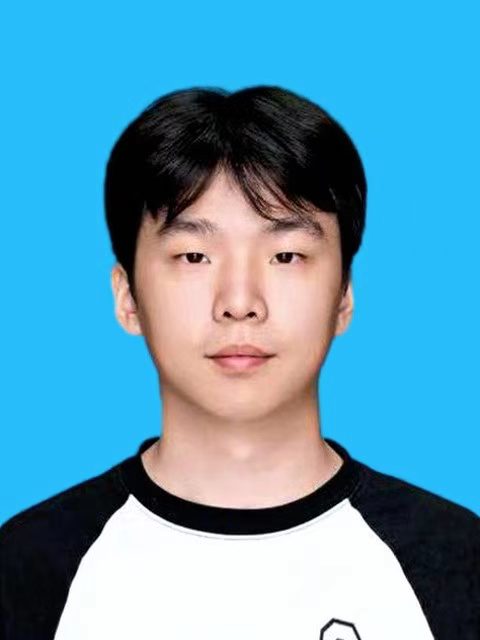}}]{Feng Zhu} is currently a researcher at GoPlus Security. His research interests mainly focus on blockchain and risk management. For more information, kindly visit https://github.com/zhufeng7.
\end{IEEEbiography}

\vspace{-33pt}

\begin{IEEEbiography}[{\includegraphics[width=1in,height=1.25in,clip,keepaspectratio]{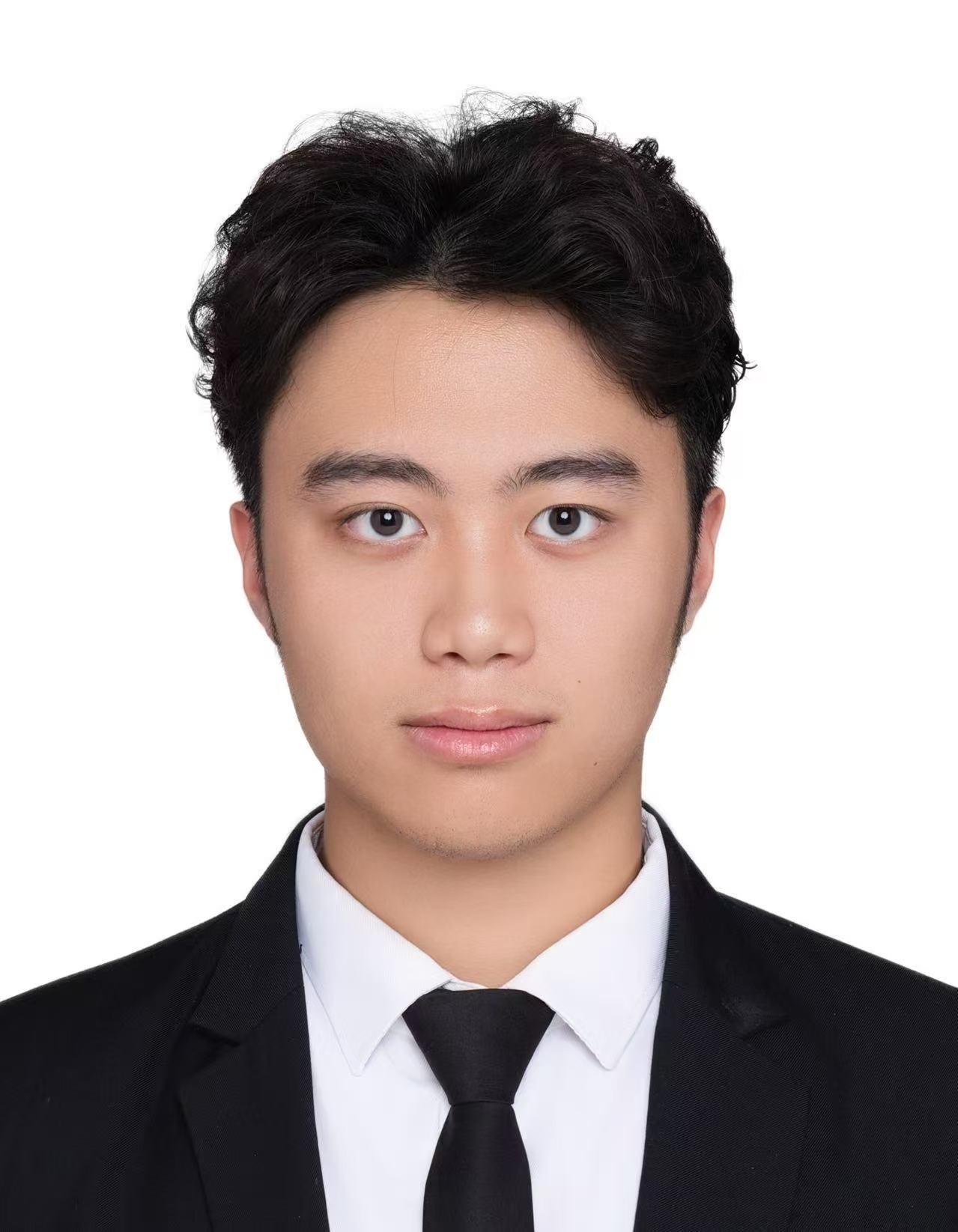}}]{Zhengyu Sun} is currently a master's student in the College of Computing and Data Science, Nanyang Technological University. His research interests mainly focus on human-computer interaction and human-AI collaboration in data visualization development. He received his bachelor's degree in Software Engineering from Shandong University in 2025.
\end{IEEEbiography}

\vspace{-33pt}

\begin{IEEEbiography}[{\includegraphics[width=1in,height=1.25in,clip,keepaspectratio]{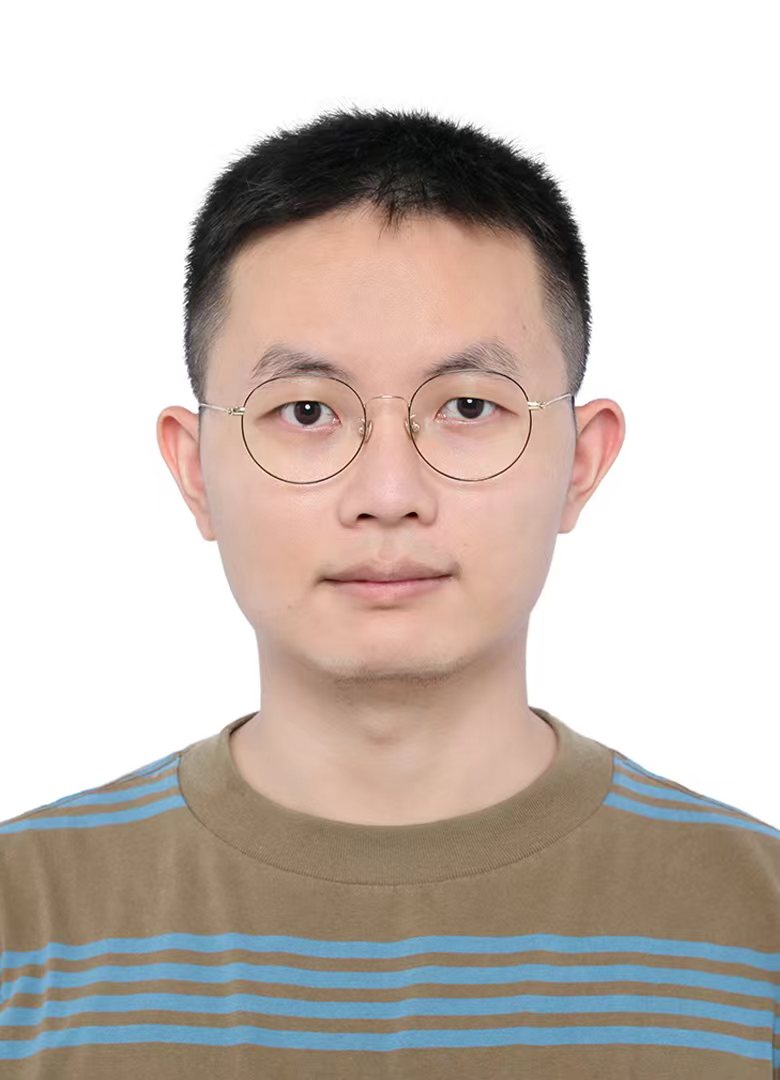}}]{Can Liu} is a research fellow in the College of Computing and Data Science, Nanyang Technological University, Singapore.
He received his Ph.D. in Intelligent Science (2023), B.Sc. in Computer Science (2018), and B.Ec in Economics (2018) degrees from Peking University.
His research interests lie in the field of deep learning-driven visualization, especially intelligent interaction for visualization. For more details, please visit \url{https://liucan.me}.
\end{IEEEbiography}

\vspace{-33pt}

\begin{IEEEbiography}[{\includegraphics[width=1in,height=1.25in,clip,keepaspectratio]{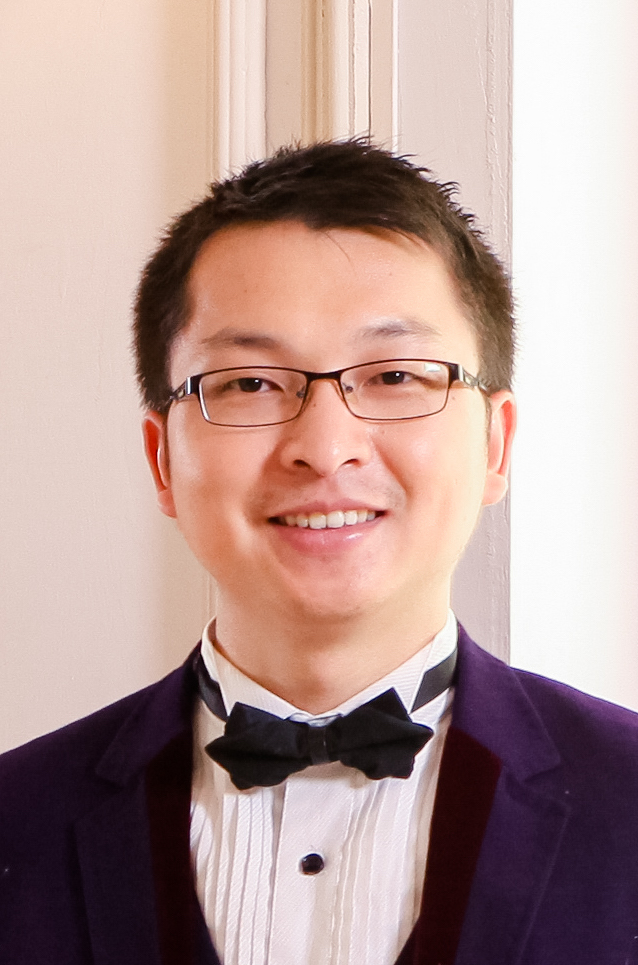}}]{Yong Wang} is currently an assistant professor in the College of Computing and Data Science, Nanyang Technological University. His research interests include data visualization, HCI, and human-AI collaboration. He obtained his Ph.D. in Computer Science from Hong Kong University of Science and Technology. He received his B.E. and M.E. from Harbin Institute of Technology and Huazhong University of Science and Technology, respectively. For more details, please refer to \url{http://yong-wang.org}.
\end{IEEEbiography}

\clearpage
\onecolumn
\appendices
\input{source/7_appendix}

\end{document}

%% file: source/1_intro.tex
\section{Introduction} \label{sec:intro}
\IEEEPARstart{M}{arket} manipulation in cryptocurrency markets remains a persistent concern due to limited regulatory oversight~\cite{xiong2023market, eigelshoven2021cryptocurrency, chainalysis2025marketmanipulation}. 
Prior studies and reports have documented diverse manipulation behaviors, including wash trading, pump-and-dump schemes, and coordinated trading~\cite{eigelshoven2021cryptocurrency, pennec2021wash, pump_dump_survey}.
These behaviors can mislead market participants and result in substantial financial losses. 
In 2024, suspected wash trading on popular blockchains accounted for up to \$2.57 billion in trading volume, and approximately 4\% of newly launched cryptocurrencies exhibited patterns associated with pump-and-dump schemes~\cite{chainalysis2025cryptocrime}. 
In this work, we focus on trade-based manipulation, 
as on-chain trading activities can provide direct evidence for inferring suspicious trading behaviors~\cite{xiong2023market}. 
This assessment is important for helping investors evaluate potential risks before making investment decisions.

Prior studies have mostly treated market manipulation as a retrospective detection problem and identified suspicious activities using statistical measures~\cite{cong2023crypto,sila2025determinants,kamps2018moon}, heuristic rules~\cite{victor2021detecting,wang2025wash,tovsic2025beyond}, or machine learning techniques~\cite{nghiem2021detecting,wu2025profit}.
These methods are useful for detecting known patterns, but provide limited support for revealing emerging or adaptive manipulation strategies and explaining why an activity is suspicious.
In practice, investors assess potential manipulation risks by combining domain knowledge with evidence from holder structures, trading behaviors, and market dynamics.
Visual analytics approaches can support such multi-perspective evidence exploration, but existing studies and tools often focus on a specific manipulation type, such as wash trading~\cite{wen2023nftdisk}, or a single evidence source, such as token distribution~\cite{bubblemaps2026}.
Investors still lack flexible support for exploring diverse manipulation strategies and cross-checking heterogeneous evidence, and require substantial manual effort for comprehensive risk analysis due to the large exploration space of possible suspicious features, holder behaviors, and market signals.

Based on interviews with four cryptocurrency practitioners, we identify three key challenges (\textbf{C1--C3}) for visual analytics of cryptocurrency manipulation risk.
\textbf{C1. Difficulty in flexible evidence modeling and coordinated visualization.}
On-chain transaction data is large-scale, and manipulation strategies are diverse and complex.
Investors may interpret address relationships, suspicious behaviors, and risk implications differently depending on their domain knowledge and analysis goals, which makes it difficult to define a fixed evidence model for manipulation-risk analysis.
Moreover, risk assessment requires joint reasoning over heterogeneous signals, including address relationships, suspicious behaviors, token holdings, and price dynamics.
It is hard to flexibly model such evidence and coordinate it through interpretable visualizations for different analysis needs.
\textbf{C2. Excessive manual effort during visual analytics.}
Visual analytics is often iterative and hypothesis-driven~\cite{sacha2014knowledge}; in cryptocurrency manipulation analysis, each iteration may involve massive transaction records and complex address interactions.
Investors repeatedly inspect transaction behaviors, compare holder activities, and relate suspicious patterns to market changes across multiple views and analysis stages.
These operations make the analysis time-consuming and laborious.
It remains challenging to infer users' analytical intent and emerging hypotheses, surface relevant evidence, and reduce repetitive manual hypothesis-evidence iteration without disrupting user-driven exploration.
\textbf{C3. Intrinsic needs of rigorous evidence-based analysis.}
During visual analytics, users often derive findings by combining visually observed patterns with their domain knowledge.
However, manual analysis is prone to omissions, and users may overlook evidence that supports, contradicts, or refines an emerging hypothesis.
They may also focus on salient visual patterns during exploration while paying less attention to statistical patterns in the underlying data.
This can lead to incomplete or weakly supported interpretations of potential manipulation risks.
It remains challenging to help users systematically examine their findings, consider both supporting and contradictory evidence, and develop rigorous evidence-based conclusions.

To address these challenges, we propose \techName{}, an LLM--assisted visual analytics system for effectively analyzing cryptocurrency manipulation risks. We first conducted a preliminary study with four cryptocurrency practitioners to derive design requirements for manipulation-risk visual analysis and LLM-assisted reasoning support.
Guided by these requirements, we designed and implemented a \textit{base visual analytics system} for flexible and interpretable manipulation-risk analysis (\textbf{C1}), including coordinated views for token distributions and holder relationships, price dynamics and rule-flagged suspicious trading patterns, detailed transaction behaviors, and user-controlled data modeling and presentation.
Building on this system, we further developed a \textit{human--LLM collaborative visual analytics framework} grounded in the knowledge generation model for visual analytics~\cite{sacha2014knowledge}.
Rather than acting as a reactive assistant, the framework positions the LLM as a co-analyst by connecting a hypothesis-driven human workflow with a context-aware LLM analysis workflow.
It infers users' analytical intent and emerging hypotheses from interactions, plans follow-up analyses, and returns visual, statistical, and synthesized findings to improve analysis efficiency (\textbf{C2}) and analytical rigor (\textbf{C3}).
For evaluation, we conducted two case studies and a user study with 12 participants. 
The results suggest that \techName{} helps users flexibly examine manipulation evidence, reduces repeated evidence seeking, and organizes findings around users' hypotheses for more rigorous risk assessment.
Our main contributions are as follows:
\begin{itemize}
    \item We derive key challenges and design requirements for visual analytics of cryptocurrency manipulation risk through interviews with four cryptocurrency practitioners.
    \item We propose \techName{}, an LLM--assisted visual analytics system that integrates coordinated visualizations with a human--LLM collaborative visual analytics framework for flexible and rigorous manipulation-risk analysis.
    \item We evaluate \techName{} through two case studies and a user study with 12 participants, showing its usefulness for manipulation-risk assessment and organizing evidence around users' hypotheses.
\end{itemize}

%% file: source/2_related_work.tex
\section{Related Work}
This work is related to prior research on \textit{cryptocurrency market manipulation detection}, \textit{visualization for cryptocurrency fraud analysis}, and \textit{LLM-assisted visual analytics}.

\subsection{Cryptocurrency Market Manipulation Detection}
Existing research on cryptocurrency market manipulation detection can be broadly grouped into statistical, machine learning (ML)-based, and heuristic methods.
\textbf{\textit{Statistical methods}} flag deviations from expected market behavior using distributional, temporal, or relational statistics. For example, prior work detects wash trading through abnormal trade-size patterns or Benford-law deviations, and pump-and-dump schemes through sudden price jumps and volume surges~\cite{cong2023crypto,sila2025determinants,kamps2018moon}. These methods support market-level screening at scale, but they can be difficult to interpret, prone to false positives, and limited in capturing strategic behaviors~\cite{vivcivc2022application}.
\textbf{\textit{ML-based methods}} formulate manipulation detection as classification or anomaly detection over heterogeneous signals. Prior work combines transaction, social-media, order-book, and market signals for pump-and-dump detection~\cite{nghiem2021detecting,fantazzini2023detecting,bolz2024machine}, while recent studies use temporal graph neural networks to detect manipulation in decentralized finance markets~\cite{wu2025profit}. These methods can model complex dependencies, but often require labeled data, generalize poorly to evolving strategies, and provide limited interpretability.
\textbf{\textit{Heuristic methods}} operationalize known manipulation patterns through predefined rules. Existing studies have used such rules to detect wash trading by capturing network-based evidence~\cite{victor2021detecting,wang2025wash,tovsic2025beyond}. These approaches are interpretable, but their utility depends on hand-crafted thresholds and assumptions, and flagged behaviors still require manual inspection.
Overall, existing methods provide useful screening signals but offer limited support for flexible interpretation when ground truth is scarce, evidence is partial, and analysts need to relate suspicious behaviors to holder structures and market dynamics.

\subsection{Visualization for Cryptocurrency Fraud Analysis}
Visualization has been widely used for explainable cryptocurrency fraud analysis~\cite{tovanich2019visualization,mafrur2025blockchain}, including wash trading~\cite{wen2023nftdisk}, money laundering~\cite{bistarelli2017go,wen2025envisage}, and Ponzi schemes~\cite{wen2023code,wen2024ponzilens+}. Existing studies can be broadly categorized as code-based or transaction-based approaches. Code-based approaches reveal suspicious features in smart contract code that supports decentralized cryptocurrency trading, which is not our focus. Transaction-based approaches visualize value flows~\cite{ahmed2018tendrils,xia2020supoolvisor,di2015bitconeview,bistarelli2017go,wen2025envisage} and relationships among addresses, clusters, or exchanges~\cite{sun2019bitvis,wen2023nftdisk,sun2022bitanalysis,kinkeldey2017bitconduite,yue2018bitextract} to support tasks such as stolen-coin tracing, token-mixing inspection, money-laundering analysis, wash-trading detection, and high-frequency trading analysis~\cite{mcginn2016visualizing}.
Although these studies demonstrate the value of visualization for cryptocurrency fraud analysis, they provide limited support for manipulation-risk assessment, which requires jointly examining holder relationships, trading behaviors, suspicious-pattern evidence, and market dynamics.
Our work follows transaction-based approaches by using coordinated visualizations to support flexible risk assessment, and further integrates a human--LLM collaborative framework to reduce manual effort and strengthen evidence-based analysis.

\subsection{LLM-assisted Visual Analytics}
Recent advances in LLMs have motivated growing research on supporting visual analytics with language-based and agentic assistance, including data transformation, visualization generation, insight discovery, and analytical reasoning~\cite{agarwal2025review,ye2024generative,hutchinson2024llm,wang2023data,tian2024chartgpt}.
A related line of work studies natural-language and conversational interfaces for visual analysis, from querying visualizations and generating analytic specifications to maintaining multi-turn conversational context and visualizing analysis provenance~\cite{shen2023nli,setlur2016eviza,narechania2021nl4dv,mitra2022conversational,qu2026toa}.
More closely related to our work, recent systems use LLMs to support analytical workflows, including task planning and execution, automated exploration, onboarding and summarization, and proactive interaction-aware support~\cite{zhao2024lightva,ma2023insightpilot,zhao2024leva,zhao2025proactiveva}. These studies show the value of embedding LLMs into visual analysis workflows. In contrast, our work focuses on domain-specific manipulation-risk analysis, where the LLM does not only answer requests or suggest next steps, but infers users' analytical intent and emerging hypotheses from visual interactions and annotations, analyzes transaction data accordingly, and organizes findings that support, contradict, or refine users' reasoning.

%% file: source/3_before_method.tex
\section{Background}
This section introduces the domain concepts needed to understand our manipulation-risk analysis.

\textbf{On-chain Transactions and Holders.}
Blockchain records digital asset activities through on-chain accounts, making many transfers and trades directly observable from transaction records~\cite{nakamoto2008bitcoin,meiklejohn2013fistful}. In this paper, \textit{cryptocurrency} refers to the broader market domain, while \textit{token} refers to the specific on-chain asset under analysis. A \textit{holder} refers to a token-holding address.
Cryptocurrency trading can occur on centralized exchanges or through on-chain transactions on decentralized exchanges (DEXs)~\cite{xu2023sokdex,werner2022sokdefi}. On DEXs, trades are typically executed through liquidity pools, so both market trades and token transfers can be examined from on-chain data. 
Since our analysis is transaction-driven and focuses on evidence that can be examined through interactive visual analysis, we focus on on-chain trading activities on DEXs.

\textbf{Entity Detection in Cryptocurrency.}
Cryptocurrency users interact through on-chain addresses, but a single real-world participant may control multiple addresses to distribute holdings, split transactions, or coordinate operations~\cite{meiklejohn2013fistful,jourdan2018characterizing}. This issue is especially important on blockchains with more complex account models, such as Solana~\cite{solana2026accounts}, where address-level records do not directly correspond to behavioral entities. Entity detection~\cite{moser2022resurrecting} therefore aims to group addresses that are likely controlled by the same participant or group, enabling analysis of holding structure, inter-holder relationships, and trading behavior at a more meaningful level.
Prior studies infer related addresses from fund-flow relations, behavioral similarity, or graph structure~\cite{meiklejohn2013fistful,he2022bitcoin,tovanich2023fingerprinting,payette2017characterizing,victor2020ethereum,zhou2022behavior}. However, entity boundaries are often uncertain because ground truth is limited, and different analysis tasks may require different grouping assumptions. For manipulation-risk analysis, entity detection should therefore support task-specific, user-adjustable modeling rather than a single, fixed partition.

\textbf{Market Manipulation in Cryptocurrency.}
Manipulation is a major risk in cryptocurrency markets~\cite{xiong2023market}. Low barriers to participation, inexpensive address creation, continuous trading, and relatively limited regulation can make abnormal coordinated behaviors difficult to assess~\cite{gandal2018economics,hamrick2021pump,la2023doge,wurster2023defiranger}. Among the manipulation categories discussed in prior work~\cite{eigelshoven2021cryptocurrency}, we focus on trade-based manipulation because it is most directly reflected in on-chain trading records, fund flows, and transaction sequences~\cite{cong2023crypto,pennec2021wash,victor2021detecting}.
Following prior studies, we operationalize trade-based manipulation through two behavior types that can be examined from on-chain evidence: wash trading and coordinated same-direction trading. Wash trading refers to superficial buy-and-sell transactions conducted among the same entity or colluding entities to create artificial trading activity without genuine risk transfer. Coordinated same-direction trading refers to one or more related entities repeatedly trading in the same direction within a short time window, creating one-sided market signals that may influence other traders. In this work, these patterns are treated as suspicious evidence for risk assessment rather than definitive proof of manipulative intent. This framing motivates a joint examination of holder relationships, detailed behaviors, and temporal market dynamics.

\section{Informing the Design}
This section describes our preliminary study with four cryptocurrency practitioners and the derived design requirements.

\subsection{Preliminary Study}
The participants and procedures are detailed below:

\textbf{Participants:}
We recruited four participants from an online cryptocurrency investment community. All had substantial cryptocurrency trading experience (2--5 years). P1 and P2 were professionals in cryptocurrency-related companies: P1 led a blockchain security service group, and P2 worked on suspicious address detection as a programmer. P3 and P4 were individual investors; P3 had experience in on-chain data analysis, while P4 used online information platforms to analyze market trends and inform investment decisions.

\textbf{Procedures:}
The study consisted of two sessions. 
In the first session, four participants joined an interview lasting about 30 minutes. They described their understanding of cryptocurrency market manipulation, their typical analysis practices, and the challenges they encountered in manipulation-risk analysis. 
Based on their feedback, we identified three key challenges (\textbf{C1--C3}): supporting flexible and interpretable manipulation-risk analysis (\textbf{C1}), reducing repetitive manual effort (\textbf{C2}), and strengthening evidence-based analysis (\textbf{C3}). Accordingly, we derived an initial set of design requirements for both the visual analytics system and the LLM-assisted analysis support. In the second session, we invited the same participants to review and refine these requirements. The final design requirements are presented in Section~\ref{sec:requirements}.

\subsection{Design Requirements} \label{sec:requirements}
We organize the design requirements (R1--R6), derived from our preliminary study, into two aspects: \textit{Visual Analytics System} and \textit{LLM Assistance}.
Participants described manipulation-risk assessment from two complementary perspectives: holders' capability to influence the market and their likelihood of manipulation. Capability refers to substantial token ownership, whereas likelihood concerns suspicious relationships or prior manipulation behaviors. 
R1--R4 describe how the visual analytics system supports flexible manipulation-evidence modeling and coordinated visualization (C1).

\textbf{R1. Reveal token distribution among holders.}
Token distribution directly indicates holders' capability to influence the market, as holders with substantial ownership may exert stronger effects on price movements and liquidity. Existing tools, such as Bubblemap~\cite{bubblemaps2026}, commonly support top-holder inspection. However, effective analysis requires more flexible support: users should be able to adjust top-holder thresholds, and accounts likely controlled by the same entity should be grouped to reveal actual ownership concentration.

\textbf{R2. Analyze customizable multi-level holder relationships.}
Holder relationships provide important clues for assessing potential collusion risks. Such relationships may arise from direct token transfers, common funding sources, or similar trading patterns. Because users may hold different assumptions about which relationships are meaningful, the system should allow them to flexibly define and explore multi-level relationships and examine these relationships together with token ownership to assess manipulation risks.

\textbf{R3. Characterize suspicious trading patterns and market impacts.}
Trade-based manipulation can take multiple forms, such as wash trading and coordinated trading. These behaviors often lack clear ground truth and are interwoven with normal trading activities, requiring users to interpret them based on expertise and assumptions. The system should therefore allow users to specify manipulation-like patterns, flag suspicious trading activities under user-defined rules, and present their market impacts intuitively to support risk analysis.

\textbf{R4. Inspect detailed holder behaviors.}
Because inferred holder relationships and suspicious-pattern flagging rules are inherently subjective, users need to inspect actual holder behaviors to validate findings and uncover finer-grained insights. The system should support detailed inspection of selected holders and related accounts, including their trading and transfer activities, balance changes, and profit or loss patterns.

R5--R6 describe how LLM-assisted visual analytics reduces repetitive hypothesis-evidence iteration (C2) and supports rigorous evidence-based reasoning (C3).

\textbf{R5. Infer users' analytical intent to reduce repetitive hypothesis-evidence iteration.}
Manipulation risk analysis often involves repeated transitions between hypothesis formation and evidence seeking. After observing suspicious patterns, users need to inspect additional views, adjust analysis parameters, and retrieve further behavioral or statistical evidence to verify or refine their hypotheses. Such iterative operations are labor-intensive and may interrupt the flow of exploration. Therefore, the system should infer users' analytical intent and emerging hypotheses from their interactions and current visual context, and proactively surface relevant findings to reduce repetitive effort in hypothesis generation and validation.

\textbf{R6. Enable collaborative and evidence-based reasoning between users and LLMs.}
Manipulation-risk analysis requires more than isolated suggestions or final conclusions. Users need assistance that understands their reasoning process, collaborates with them during exploration, and strengthens their interpretations with evidence. The system should therefore support interaction-aware human--LLM collaboration, where the LLM contributes relevant evidence and alternative perspectives that support, contradict, or refine users' hypotheses, helping them form more rigorous conclusions.

%% file: source/4_method.tex

\section{\techName{}}
Guided by the design requirements, we develop \techName{} as a system that consists of a base visual analytics system (R1--R4) and a human-LLM collaborative visual analytics framework (R5--R6).
Details are described as follows.

\subsection{Base Visual Analytics System}
The base visual analytics system of \techName{} is introduced through its data description, user-customizable processing modules, and base visual analytics interface. In particular, the data description and processing modules support flexible evidence modeling (R1--R3), while the interface coordinates the resulting views for interactive analysis (R1--R4).

\subsubsection{Data Description}
\label{sec:datadesc}
We selected two representative tokens, ACT~\cite{actbinance2024} and PNUT~\cite{pnutcoingecko2026}, as example datasets. Both tokens have high trading activity and strong price volatility during their rapid growth periods, making them suitable cases for studying potential manipulation risks. We collected their on-chain transactions from Dune Analytics~\cite{dune2026}, covering the period from token launch to exchange listing. As our focus is on visual analytics and LLM-assisted analysis, real-time data updates are left for future work.
Table~\ref{tab:dataset_statistics} shows the statistics of the collected dataset, including the time span (UTC), number of transactions (\#Txns), transaction volume (USD / token amount), and number of unique addresses (\#Addr.).
We further identified exchange and liquidity-pool addresses, incorporated Solscan labels for known accounts, and generated multi-resolution price dynamics. Transfer records support balance and relationship analysis, while trade records support suspicious-pattern detection, behavioral inspection, and market-impact analysis throughout the system.

\begin{table}[htbp]
\centering
\caption{Statistics of Trades and Transfers for ACT and PNUT.}
\label{tab:dataset_statistics}
\resizebox{\columnwidth}{!}{%
\begin{tabular}{lcccc}
\toprule
\multirow{2}{*}{\textbf{Metric}} 
& \multicolumn{2}{c}{\textbf{ACT}} 
& \multicolumn{2}{c}{\textbf{PNUT}} \\
\cmidrule(lr){2-3} \cmidrule(lr){4-5}
& \textbf{Trades} & \textbf{Transfers} & \textbf{Trades} & \textbf{Transfers} \\
\midrule

Start (UTC) 
& 2024-10-19 11:27 & 2024-10-19 11:27 
& 2024-10-31 14:21 & 2024-10-31 14:21 \\

End (UTC)   
& 2024-11-09 23:59 & 2024-11-09 23:59 
& 2024-11-09 23:59 & 2024-11-09 23:59 \\

\# Txns     
& 989,606 & 1,246,803 
& 1,099,197 & 1,667,411 \\

Volume      
& \$630.65M / 37.42B & 49.73B 
& \$947.57M / 48.57B & 60.99B \\

\# Addr.    
& 63,056 & 69,986 
& 98,707 & 109,782 \\

\bottomrule
\end{tabular}%
}
\end{table}

\subsubsection{Data Processing Module for User-Customized Configuration}\label{sec:processing}
To support user-customizable data modeling (R1--R3), our system provides dynamic computation services that adapt the analytical results to users' selected analysis time points, top-holder thresholds, relationship definitions, and suspicious-pattern flagging rules.
Details are described as follows:

\textbf{Token Distribution Processing:}
To support flexible and concise token distribution analysis (R1), we allow users to select balance snapshots at any timestamp and specify a threshold to filter the \textit{top holders} for detailed inspection, while aggregating the remaining holders into an \textit{Others} category. Specifically, when the threshold is set to 0.3, we select the top holders whose cumulative balances, ranked in descending order, account for 30\% of the total token holdings. This design allows users to focus on the most influential holders while preserving a concise overview of the overall distribution. In addition, we identify holders that directly transact with top holders as \textit{related holders}, preserving their local transaction context for subsequent relationship and behavior analysis.

\textbf{Entity and Relationship Detection:}
To support flexible analysis of multi-level holder relationships (R2), we define a set of parameterized rules for identifying potential entities and inter-holder relationships. Users can customize the parameters of these rules to reflect the types of connections they consider meaningful in different analytical scenarios. 
The same rule framework is applied to both entity detection and relationship detection, with stricter settings used to identify holders likely controlled by the same entity and relatively looser settings used to capture potential associations among holders. This design allows users to adapt relationship modeling to their analytical preferences.
The rules are grouped into three categories:
\textbf{\textit{Network-based rules}} capture fund-flow associations through transaction networks, such as direct transfers, shared funding sources, and common counterparties;
\textbf{\textit{Similarity-based rules}} identify holders with similar behavioral patterns from action, balance, and earnings sequences under user-defined temporal granularities and similarity thresholds;
\textbf{\textit{Suspicious-pattern-based rules}} connect holders through shared suspicious trading activities flagged under user-defined settings.

\begin{figure*}[t]
  \centering
  \includegraphics[width=\linewidth]{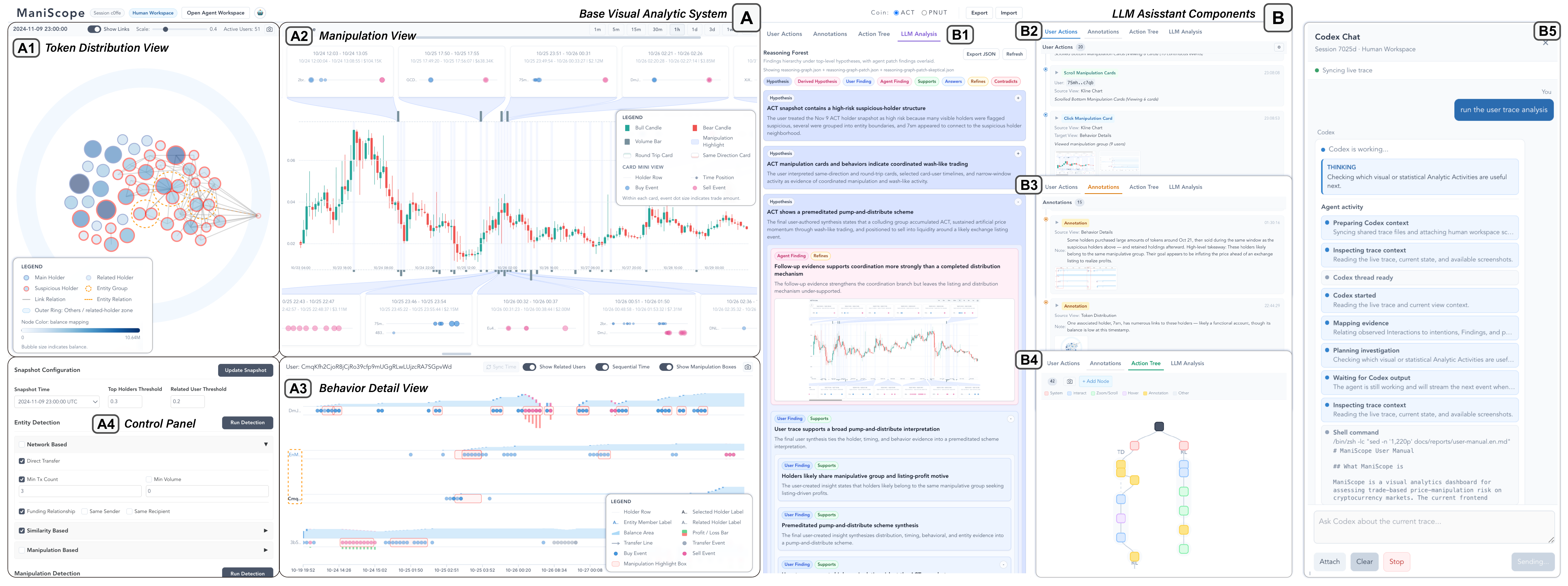}
  \caption{The \techName{} interface. (A) The base visual analytics system consists of the Token Distribution View (A1), Manipulation View (A2), Behavior Detail View (A3), and Control Panel (A4). (B) The human-LLM collaborative interface augments the base system with an LLM analysis panel (B1), user action and annotation histories (B2--B3), an action tree for tracking the analytical process (B4), and a chat panel for interactive requests and execution feedback (B5).}
  \label{fig:interface}
\end{figure*}

\textbf{Suspicious Trading Pattern Flagging:} 
To help users inspect manipulation-like activities within the available transaction history (R3), we define parameterized rules for flagging suspicious trading patterns, focusing on potential \textit{wash trading} and \textit{coordinated same-direction trading}. These rules can be applied either to individual holders or aggregated at the entity level using the results from entity detection. This allows users to examine suspicious patterns that may appear at the address level or emerge only after related addresses are grouped. The flagging process is driven by two primary rule sets:
\textbf{\textit{Round Trip Rule}} flags potential wash-trading patterns by identifying temporally concentrated buy-sell sequences with near-zero net position change and earnings under user-defined thresholds;
\textbf{\textit{Same Direction Rule}} flags potential coordinated same-direction patterns by identifying temporally concentrated sequences of repeated buying or repeated selling under user-defined sequence and tolerance settings.

\subsubsection{Base Visual Analytics Interface}

The base visual analytics interface of \techName{} (Fig.~\ref{fig:interface}\,\captionID{A}) consists of four coordinated components: the \textit{Token Distribution View}, \textit{Manipulation View}, \textit{Behavior Detail View}, and a \textit{Control Panel}.

\textbf{\textit{Token Distribution View}} (Fig.~\ref{fig:interface},\captionID{A1}) is an enhanced node-link diagram that jointly depicts token distributions among top holders (R1) and relationships between them (R2). \textit{Top holders}, filtered based on a user-defined threshold (Sec.~\ref{sec:processing}), are represented as nodes within a circular ring, where node size and a blue gradient encode token balances; larger and darker nodes indicate greater holdings. The token holdings of all remaining holders are aggregated into an \textit{Others} category, represented by the outer ring whose area is scaled consistently with node size.
To illustrate holder relationships, we additionally visualize \textit{related holders} (Sec.~\ref{sec:processing}) as nodes placed within the outer ring, using the same encoding as top holders. Top holders identified as belonging to the same entity are grouped within an orange dashed circle and spatially packed together, while related holders are connected to the dashed circle via orange dashed links. Grey links between holders represent relationships detected based on user-defined rules, whereas related holders are included based on direct transaction proximity. Holder nodes are arranged using a force-directed layout with collision avoidance: linked nodes are positioned cohesively, while top holders remain within the inner ring and related holders within the outer ring.
Suspicious nodes associated with previously flagged suspicious trading patterns are highlighted with a red stroke. Users can hover over nodes to view detailed information, including wallet address, token balance, and reasons for being flagged. Overall, the \textit{Token Distribution View} helps users examine token concentration, holder relationships, entity-level connections, and prior suspicious behaviors for manipulation-risk assessment.

\begin{figure*}[t]
  \centering
  \includegraphics[width=\linewidth]{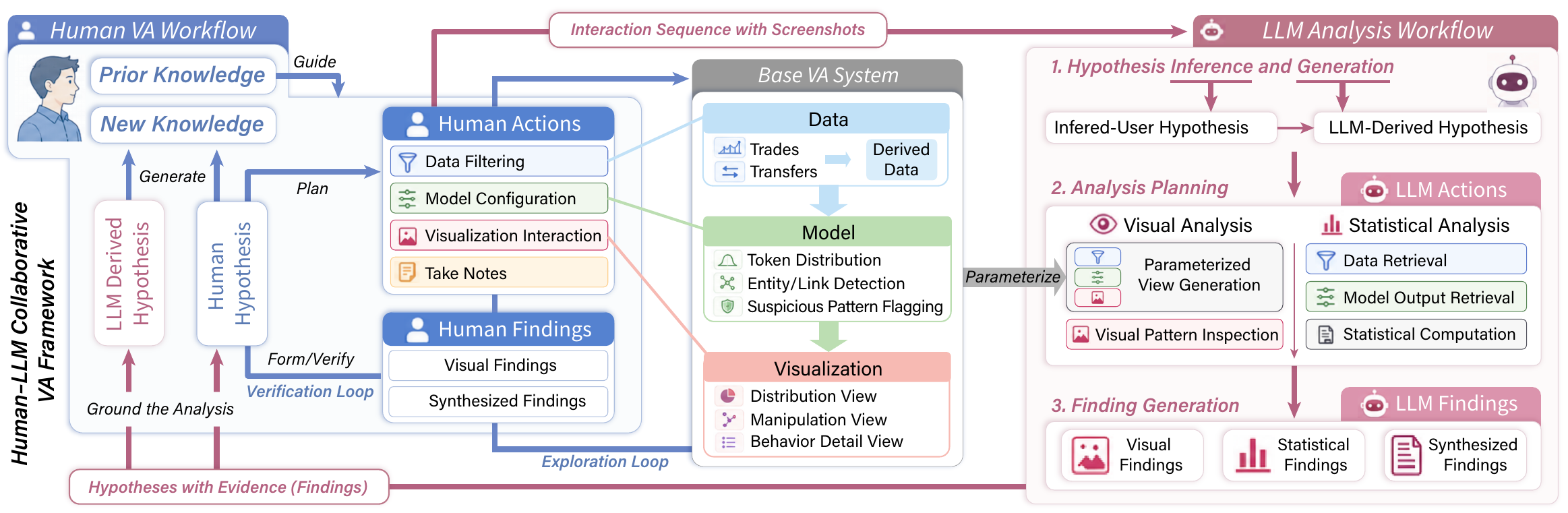}
  \caption{Overview of ManiScope's human-LLM collaborative visual analytics framework. The framework combines a shared data-model-visualization substrate, a hypothesis-driven human visual analytics workflow, and a context-aware LLM analysis workflow that supports bottom-up hypothesis inference, top-down analysis planning, and evidence-based finding generation for hypothesis verification.}
  \label{fig:framework}
\end{figure*}

\textbf{\textit{Manipulation View}} (Fig.~\ref{fig:interface},\captionID{A2}) shows price dynamics together with rule-flagged suspicious trading patterns over time (R3). We use a candlestick chart to visualize price movements under user-selected temporal granularities (e.g., 5 min, 1 h, and 1 d). Each candlestick encodes the open, close, high, and low prices, with color indicating whether the price rises or falls.
Above and below the candlestick chart, grey bar charts show the number of flagged \textit{round-trip} and \textit{same-direction} patterns in each interval, with corresponding intervals highlighted by light blue bands. Detailed pattern cards present the suspicious sequences detected in each interval: the y-axis lists involved participants, and the x-axis shows the action sequence, where buy and sell actions are encoded as blue and pink circles, respectively.
Users can zoom in and out to navigate intervals of interest. The pattern cards can be scrolled independently while remaining visually linked to their corresponding intervals through light blue bands. This view helps users examine suspicious trading patterns in relation to their temporal context and price impacts (R3).

\textit{\textbf{Behavior Detail View}} (Fig.~\ref{fig:interface}\,\captionID{A3}) presents detailed behaviors of holders associated with a user-selected holder, as explored from the \textit{Token Distribution View}.
The y-axis lists the selected holder and all holders that have entity or link relationships with it, with the selected holder and those belonging to the same entity positioned at the center.
The x-axis represents a timeline, illustrating holder behaviors across three aspects: actions, balance, and earnings.
We encode transaction actions using colored circles, where blue, pink, and grey represent buy, sell, and transfer actions of the target token, respectively.
Holders' balances over time are visualized using a blue area chart, with blue and pink bars overlaid to indicate balance changes resulting from buy and sell actions.
Earnings are encoded using bars positioned beneath the action markers, where red and green represent losses and gains, respectively.
Additionally, red bounding boxes highlight action sequences and time periods flagged as suspicious based on user-defined rules specified in the control panel.
Users can zoom in and out to examine behavioral details across time periods of interest.
This view enables users to inspect flagged patterns and validate inferred relationships through detailed behavioral evidence (R4).

\textbf{\textit{Control Panel}} (Fig.~\ref{fig:interface},\captionID{A4}) supports configurable data processing and evidence modeling (R1--R3). It includes \textit{snapshot configuration} for selecting snapshot time and holder-filtering thresholds, \textit{entity detection configuration} and \textit{link detection configuration} for defining entity and relationship rules based on network-, similarity-, or suspicious-pattern-based criteria, and \textit{suspicious-pattern configuration} for specifying manipulation-like pattern types and flagging parameters.

\subsection{Human--LLM Collaborative Visual Analytics Framework}
\label{sec:llm_framework}

Building on the base visual analytics system, we further propose a \textit{human--LLM collaborative visual analytics framework} to reduce repetitive hypothesis-evidence iteration (R5) and strengthen evidence-based reasoning (R6).
Specifically, the framework models the \textit{human visual analytics workflow} for manipulation-risk analysis as a hypothesis-driven knowledge generation process and augments it with a \textit{context-aware LLM analysis workflow} for hypothesis inference, analysis planning, and finding generation.
Fig.~\ref{fig:framework} shows the framework overview, which is conceptually grounded in the \textit{knowledge generation model for visual analytics}~\cite{sacha2014knowledge}. 
We use this model to characterize how knowledge is produced during manipulation-risk analysis and to determine where LLM assistance should be introduced.
In particular, the base visual analytics system instantiates the model's \textit{data}, \textit{model}, and \textit{visualization} components, while the human workflow captures users' exploration through filtering, model configuration, visualization interaction, and note-taking.
Users observe initial \textit{findings} during exploration, form tentative \textit{hypotheses}, plan further \textit{actions}, and collect evidence as \textit{findings} to verify or refine these hypotheses.
This formulation captures how users gradually develop \textit{knowledge} from visual and statistical evidence.

\begin{figure*}[!t]
  \centering
  \includegraphics[width=\linewidth]{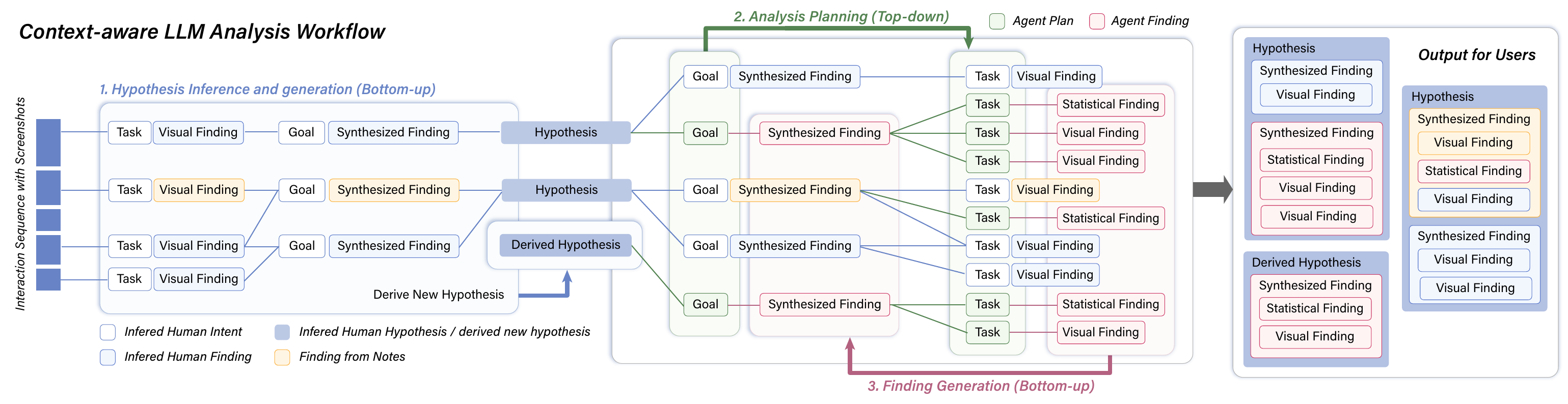}
  \caption{The LLM workflow in \techName{} consists of three stages: bottom-up hypothesis inference and generation from interaction context, top-down analysis planning for expanding the analytical hierarchy, and finding generation that produces visual, statistical, and synthesized findings for hypothesis verification.}
  \label{fig:llmworkflow}
\end{figure*}

For the \textit{context-aware LLM analysis workflow}, the LLM is embedded into the exploration and verification process as a co-analyst, rather than an independent suggestion module. 
It interprets user interactions and current visual context, conducts corresponding analyses over the shared analytical substrate, and returns evidence-based findings that help verify, refine, or extend users' hypotheses.
Fig.~\ref{fig:llmworkflow} further details this LLM workflow in three stages.
The corresponding interface components are shown in Fig.~\ref{fig:interface}\,\captionID{B}.
\techName{} provides an LLM analysis panel for inspecting generated reasoning and findings (B1), user action and annotation histories for context capture (B2--B3), an action tree for tracking the ongoing analytical process (B4), and a chat panel for interactive requests and execution feedback (B5).
The chat panel supports on-demand assistance, such as explaining visual encodings, summarizing visible data, answering questions about the current view, and suggesting possible next steps.
When users trigger LLM-assisted analysis, \techName{} uses the recorded interactions and annotations to infer hypotheses, plan follow-up analyses, and return evidence-based findings through the collaborative framework.

\textbf{Hypothesis Inference and Generation:}
To address R5, we introduce a context-aware bottom-up modeling strategy for hypothesis inference and generation. This stage converts the user interactions into explicit analytical hypotheses, while establishing a semantic basis for subsequent evidence-based reasoning. As illustrated in Fig.~\ref{fig:llmworkflow}, the agent takes users' interaction sequences with screenshots of the current visual context as input and progressively infers hypotheses from low-level interaction evidence rather than generating them in a single step.
Specifically, the modeling process is organized into three levels. At the first level, the agent interprets each interaction action in context and infers the corresponding \textit{finding} and analytical \textit{task}. At the second level, it aggregates findings from multiple related actions into middle-level \textit{synthesized findings} and infers the associated \textit{analytic goal}. At the third level, it infers the \textit{hypothesis} that the user is currently forming or attempting to verify based on the accumulated goals and findings. Beyond recovering user hypotheses, the agent can further derive new hypotheses that extend the current reasoning path. These derived hypotheses do not replace user judgment; instead, they provide additional directions for verification and help users conduct more efficient and rigorous analysis.

\textbf{Analysis Planning:}
Based on the hierarchical analytical structure inferred in the previous stage, analysis planning identifies gaps in the user's current reasoning and proposes additional directions for verification. Rather than directly producing findings from the inferred hypothesis, this stage examines the existing hierarchy of tasks, findings, synthesized findings, goals, and hypotheses (Fig.~\ref{fig:llmworkflow}) to detect unsupported claims, missing evidence, or unexplored branches. It then expands the hierarchy by introducing new \textit{goals} or executable \textit{tasks} that guide subsequent analysis.
As shown in Fig.~\ref{fig:framework}, the agent performs top-down planning from the current hypothesis and its associated goals, decomposing missing analytical support into visual and statistical operations. The visual branch parameterizes relevant views and inspects suspicious structural or temporal patterns, while the statistical branch retrieves raw data and model outputs for corresponding computations. In this way, analysis planning guides the system toward findings that can strengthen, challenge, or complete the ongoing analysis.

\textbf{Finding Generation:}
To support R6, we produce evidence from the planned tasks and organize it into forms that directly support verification. As shown in Fig.~\ref{fig:llmworkflow}, the agent first executes the tasks proposed during the analysis planning stage using either visual analytics or statistical actions. For visual analytic actions, the agent generates visualizations under the current configurations and inspects the resulting patterns to obtain \textit{visual findings}. For statistical actions, the agent retrieves the required data and model outputs and executes scripts or computations to obtain \textit{statistical findings}.
These low-level findings are then integrated according to the \textit{analytic goals} inferred in the hypothesis modeling stage. Findings associated with the same goal are synthesized into \textit{synthesized findings}, summarizing how multiple pieces of evidence relate to the current line of reasoning. The synthesized findings are finally organized around current and derived hypotheses and returned as interpretable evidence that supports or contradicts them. In this way, the system does not merely return isolated outputs, but contributes structured evidence for more rigorous human-LLM collaborative reasoning.

\subsection{System Implementation}
We implement \techName{} as a web-based prototype with a Vue front end and a Python/FastAPI backend. The backend manages processed trades and transfers, computes configurable balance snapshots, holder relationships, entities, and suspicious-pattern flags, and serves these results to the visual analytics interface. The LLM workflow uses the same analytical substrate: it records user interactions and screenshots, invokes a model-agnostic LLM backend (GPT-5.5 in this work) for hypothesis inference, planning, and finding generation, and executes visual or statistical follow-up actions when evidence is needed. Visual actions render parameterized views, whereas statistical actions run scripts over retrieved data and model outputs. The generated hypotheses, findings, and execution feedback are returned for user inspection and refinement.

%% file: source/5_evaluation.tex
\begin{figure*}[t]
  \centering
  \includegraphics[width=\linewidth]{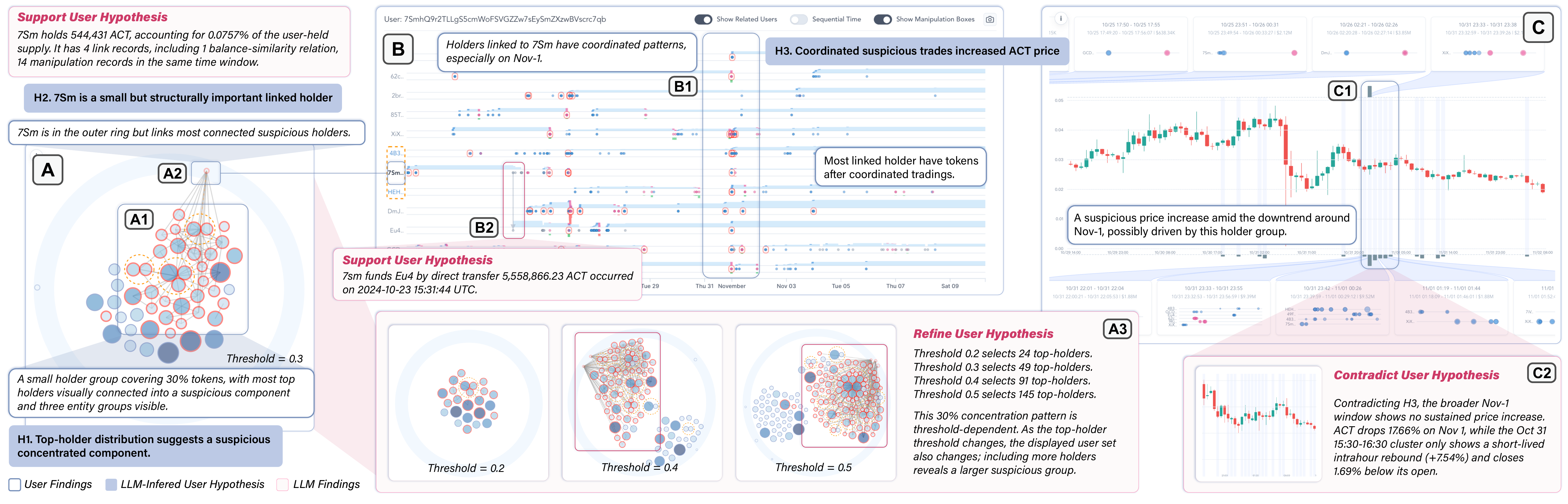}
  \caption{A case showing how LLM assistance reduced manual follow-up in ACT. P1 explored the token distribution (A), behavior details (B), and suspicious-pattern timeline (C), while the LLM inferred implicit hypotheses and returned supporting (B2), refining (A3), and contradicting findings (C2).}
  \label{fig:case_study_act}
\end{figure*}

\section{Case Studies}
To illustrate how \techName{} supports manipulation-risk analysis, we invited two experienced participants from our preliminary study, P1 and P2, to conduct case studies on ACT and PNUT. Each participant analyzed one dataset with \techName{}, and each session lasted approximately one hour.
Rather than reporting the full process, we summarize the main reasoning path from each session to show how \techName{} supports key analysis tasks (R1-R4), while reducing manual effort (R5) and strengthening evidence-based analysis (R6).

\subsection{Letting the LLM Handle Repetitive Validation}

P1 began from the \textit{Token Distribution View} with the top-holder threshold set to 0.3 (Fig.~\ref{fig:case_study_act}\captionID{A}), using it to inspect token concentration and holder relationships together (R1--R2). Based on his domain knowledge, he treated the compact holder component as suspicious and recorded notes: the holder group covered about 30\% of the token supply, most top holders were connected, and three entity groups were visible (Fig.~\ref{fig:case_study_act}\captionID{A1}). He also noticed that \textit{7sm} held a small balance but linked many suspicious holders (Fig.~\ref{fig:case_study_act}\captionID{A2}), making it suspicious. After clicking the \textit{7sm} node, P1 entered the \textit{Behavior Detail View} (Fig.~\ref{fig:case_study_act}\captionID{B}) and found coordinated patterns among holders linked to \textit{7sm}, especially around Nov. 1 (R4; Fig.~\ref{fig:case_study_act}\captionID{B1}). He then checked the same period in the \textit{Manipulation View} (Fig.~\ref{fig:case_study_act}\captionID{C}) and observed suspicious activities near a local price rebound (R3; Fig.~\ref{fig:case_study_act}\captionID{C1}). P1 then clicked a button to trigger LLM assistance and continued exploring the VA interface, while the LLM began to support the follow-up analysis:

\textbf{Sweeping parameters without repeated manual trials.}
The LLM inferred implicit hypotheses from the saved interaction records and notes: the top-holder distribution might indicate a suspicious concentrated component (H1), \textit{7sm} might be a small but structurally important connector (H2), and coordinated suspicious trades might have affected ACT's short-term price (H3). It then planned follow-up analyses to reduce the manual effort needed to validate these hypotheses. For H1, the LLM swept alternative top-holder thresholds instead of requiring P1 to repeatedly adjust parameters and compare holder distributions (Fig.~\ref{fig:case_study_act}\captionID{A3}). This parameter sweep produced a concise statistical and visual finding: the 30\% concentration pattern was threshold-dependent because the displayed holder set changed with the threshold, while nearby thresholds still revealed a larger suspicious group. This helped P1 preserve the main structural interpretation while understanding its parameter sensitivity.

\textbf{Finding more evidence without manual guidance.}
The LLM also searched for additional findings that P1 would otherwise need to collect through cross-view iteration. For H2, it summarized statistical evidence that \textit{7sm} had multiple link records, including balance-similarity and same-window suspicious-pattern records. More importantly, it organized a visual finding in the \textit{Behavior Detail View}: a direct transfer from \textit{7sm} to \textit{Eu4} (Fig.~\ref{fig:case_study_act}\captionID{B2}). P1 later noted that he had overlooked this transfer during his manual exploration, but the LLM surfaced it as relevant evidence for the connector hypothesis. For H3, the LLM checked the broader Nov. 1 window and generated a contradicting finding: ACT did not show sustained price support, and the selected suspicious cluster only overlapped with a short-lived intrahour rebound (Fig.~\ref{fig:case_study_act}\captionID{C2}). 
In this case, \techName{} kept P1's domain reasoning as the starting point, but automatically inferred implicit hypotheses and gathered supporting, refining, and contradicting evidence around them, reducing the burden of repeated manual follow-up analysis. These findings helped P1 assess ACT as showing a suspicious concentrated holder structure and localized coordination risk, while also recognizing limited evidence for price impact.

\begin{figure*}[t]
  \centering
  \includegraphics[width=\linewidth]{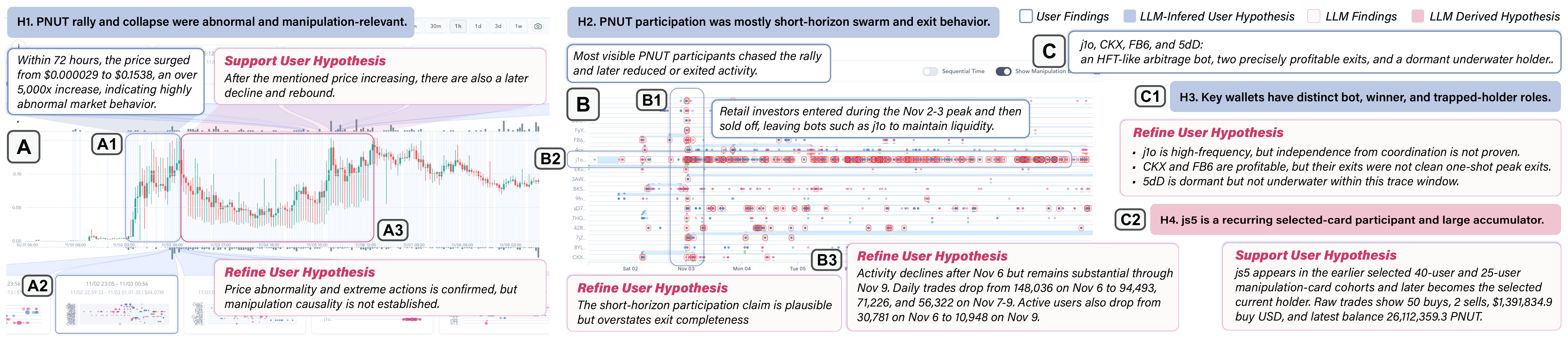}
  \caption{A case showing how LLM assistance made PNUT analysis more evidence-based. P2 explored the suspicious-pattern timeline (A), expanded a selected flagged card into behavior details (B), and inspected key wallet behaviors (C), while the LLM supported or refined inferred hypotheses (A3, B3, C1) and derived a new \textit{js5} hypothesis with supporting evidence (C2).}
  \label{fig:case_study_pnut}
\end{figure*}

\subsection{Case 2: Grounding the Analysis in Evidence}
P2 first checked the \textit{Manipulation View}, which showed an extreme rally followed by a collapse (Fig.~\ref{fig:case_study_pnut}\captionID{A}). He noted that the price increased by more than 5,000 times within 72 hours (Fig.~\ref{fig:case_study_pnut}\captionID{A1}), and treated this as an abnormal market pattern. He then clicked a flagged suspicious-pattern card around the Nov. 2--3 peak (Fig.~\ref{fig:case_study_pnut}\captionID{A2}), which expanded the \textit{Behavior Detail View} (Fig.~\ref{fig:case_study_pnut}\captionID{B}). This linked the market-level pattern to holder-level behavioral evidence. From the visible holder timelines, P2 observed that many participants entered during the peak and later reduced or exited activity (Fig.~\ref{fig:case_study_pnut}\captionID{B1}). He also continued checking the behavior details of several key holders, which are summarized in Fig.~\ref{fig:case_study_pnut}\captionID{C}: \textit{j1o} appeared high-frequency, \textit{CKX} and \textit{FB6} appeared profitable, and \textit{5dD} appeared dormant. P2 then triggered the LLM to infer his analytical focus and help verify the emerging narrative.

\textbf{Making existing hypotheses more evidence-based.}
From P2's interaction records and notes, the LLM inferred three hypotheses: PNUT's rally and collapse were worth examining for manipulation risk (H1), most visible participants showed short-horizon swarm-and-exit behavior (H2), and several wallets played distinct roles (H3). It then generated visual and statistical findings to support or refine them. For H1, the LLM supported the observed price abnormality, but refined the claim by separating abnormal market movement from established manipulation causality (Fig.~\ref{fig:case_study_pnut}\captionID{A3}). For H2, it corroborated that the short-horizon participation pattern was plausible, while statistical findings showed that the exit pattern was incomplete: activity declined after Nov. 6, but daily trades and active users remained substantial through later days (Fig.~\ref{fig:case_study_pnut}\captionID{B3}). For H3, the LLM used additional behavior evidence to qualify the role interpretation: \textit{j1o} was high-frequency, but independent bot behavior was not proven; \textit{CKX} and \textit{FB6} were profitable, but not clean one-shot peak exits; and \textit{5dD} was dormant, but not supported as underwater within the checked trace (Fig.~\ref{fig:case_study_pnut}\captionID{C1}). These findings made P2's narrative more solid without overstating what the evidence could support.

\textbf{Deriving and verifying a related hypothesis.}
Beyond refining the inferred hypotheses, the LLM also derived a related hypothesis from P2's analysis intent: \textit{js5} might be a recurrent selected-card participant and large accumulator (H4). It then searched for supporting findings through both visual and statistical analysis. The resulting evidence showed that \textit{js5} appeared in both the earlier 40-user and later 25-user selected flagged-card cohorts, became the selected current holder, had substantial buy activity, and retained a large PNUT balance under the checked configuration (Fig.~\ref{fig:case_study_pnut}\captionID{C2}). This derived hypothesis gave P2 a concrete follow-up direction that was consistent with his ongoing reasoning, while the generated findings helped verify it without requiring another long manual exploration loop. Overall, \techName{} helped P2 assess PNUT as a higher-risk case with abnormal market movement, short-horizon coordinated participation signals, and several wallets requiring further inspection, while avoiding unsupported claims about manipulation causality.

\section{User Study}
To further evaluate \techName{}, we conducted a user study with 12 experienced cryptocurrency practitioners.
This section reports the study settings and result analysis.

\subsection{Study Settings}
We designed the study around two evaluation goals: whether the base visual analytics system supports manipulation-risk analysis (G1; R1-R4), and whether the human-LLM collaborative framework reduces manual effort and strengthens evidence-based reasoning (G2; R5-R6). Each participant analyzed two token datasets (Sec.~\ref{sec:datadesc}), one with \techName{} and one with a baseline system.
\techName{} and the baseline shared the same interface, backend, and data. The baseline included a simplified reactive LLM assistant that could access the analysis state, answer user questions, and execute user-specified requests. 
Beyond these basic functions, \techName{} added the human-LLM collaborative framework, which inferred users' analytical intent and emerging hypotheses and returned visual, statistical, and synthesized findings.
The questionnaire package included one base visual analytics questionnaire for manipulation-risk analysis (G1), one post-condition LLM assistance questionnaire administered after each session for manual effort and evidence-based reasoning (G2), and one \techName{}-specific questionnaire for the usefulness of the human-LLM collaborative framework (G2).
The LLM output checklist further collected user ratings on the quality of generated hypotheses and findings (G2). Interviews collected qualitative feedback on participants' analysis processes, trust, and perceived limitations (G1-G2).

\textbf{Participants:}
We recruited 12 participants (U1-U12; aged 23-31, $M=26.17$; 9 male and 3 female) with experience in cryptocurrency investment, Web3 development, or blockchain analysis. 
All participants were familiar with on-chain transaction data and had prior cryptocurrency trading or analysis experience: six reported 3-5 years of experience, five reported 1-2 years, and one reported less than one year.
We used a counterbalanced within-subject design with two variables: system condition (baseline vs. \techName{}) and dataset (ACT vs. PNUT).
System order and dataset-condition pairing were balanced so that half of the participants started with each condition and both datasets appeared under both conditions across participants.
The detailed participant assignment and questionnaire items are provided in Appendix~\ref{app:user_study_details}.

\textbf{Procedures:}
We deployed both systems on a server, and participants accessed them on their own computers while sharing their screens through Zoom.
Each participant first completed a background questionnaire, a tutorial, and a short practice exploration.
Participants then completed two analysis sessions according to the counterbalanced assignment.
In each session, they used the assigned system to assess the assigned token's potential manipulation risk and thought aloud during the analysis.
Because manipulation-risk analysis depends on users' domain knowledge and subjective judgment, we did not design a narrowly scored task.
Each session lasted until the participant considered the task complete, with a maximum time limit of 30 minutes.
Participants completed the base visual analytics questionnaire regardless of condition after the first session, the post-condition LLM assistance questionnaire after each session, and the \techName{}-specific questionnaire and LLM output checklist after the \techName{} session.
The checklist assessed the alignment of generated hypotheses and the sufficiency and relevance of associated findings.
After both sessions, participants joined a semi-structured interview on their analysis experience, perceived benefits and limitations, and suggestions for \techName{}.
The study lasted about 1.5 hours, and participants received about 15 USD as compensation.
With participants' consent, we anonymously collected interaction logs, questionnaires, and checklist responses.

\subsection{Result Analysis}

\begin{figure}[t]
  \centering
  \includegraphics[width=\columnwidth]{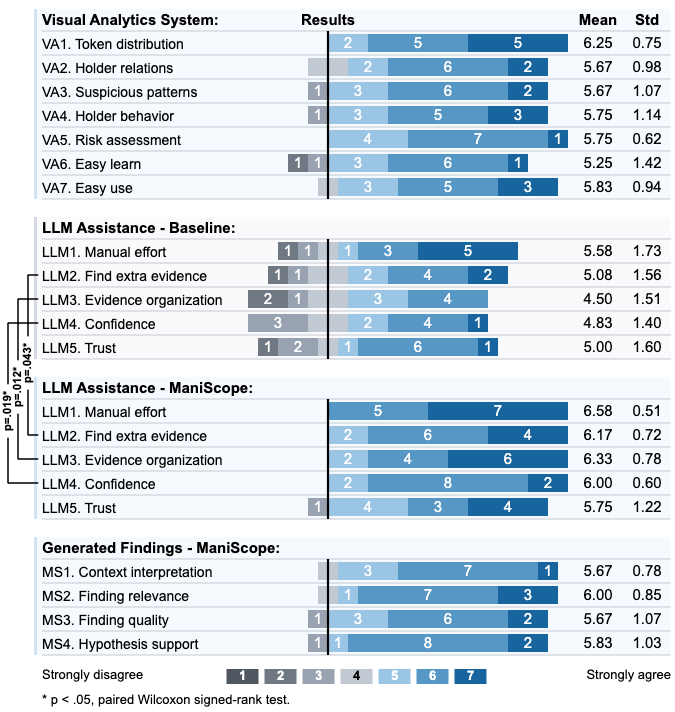}
  \caption{Questionnaire results from the user study, grouped by the base visual analytics, post-condition LLM assistance, and \techName{}-specific questionnaire. A stacked bar chart shows the distribution of 1-7 Likert ratings.}
  \label{fig:questionnaire_results}
\end{figure}

\subsubsection{Questionnaire Results}
Fig.~\ref{fig:questionnaire_results} summarizes the questionnaire results.
\textbf{Base visual analytics questionnaire.}
\techName{} received positive ratings across VA1-VA7, suggesting that participants could effectively analyze manipulation risks using the base visual analytics interface (G1). 
VA1 (M=6.25) and VA2 (M=5.67) received high ratings, indicating that the \textit{Token Distribution View} helped participants identify major holders and connected holder structures as a starting point for assessing market influence and possible coordination. 
Participants also noted that flexible filtering of major holders and their relationships supported their subjective analysis goals.
The \textit{Manipulation View} (VA3, M=5.67), \textit{Behavior Detail View} (VA4, M=5.75), and overall risk-assessment item (VA5, M=5.75) were also rated positively, suggesting that participants could move from market-level suspicious patterns to holder-level behavioral evidence and use these complementary views for manipulation-risk assessment. 
The usability results (VA6, VA7) further suggest that the system was practical for analysis, although the relatively lower ease-of-learning rating (M=5.25) indicates that its multi-view reasoning workflow benefits from tutorial guidance.

\textbf{Post-condition LLM assistance questionnaire.}
\techName{} received higher ratings than the baseline on all shared LLM-assistance items (LLM1-LLM5), with significant improvements in finding extra evidence (M=6.17 vs. 5.08, p=.043), organizing evidence (M=6.33 vs. 4.50, p=.012), and increasing evidence-supported confidence (M=6.00 vs. 4.83, p=.019).
These results suggest that \techName{} helped participants connect evidence across holder structures, suspicious-pattern timelines, detailed behaviors, and generated findings, while also extending the analysis beyond explicit user prompts.
Participants also rated \techName{} highly for reducing manual effort (M=6.58) and supporting trust in the LLM responses (M=5.75).
Together, these results support G2 by showing that the human-LLM collaborative framework reduced repeated manual checks and strengthened evidence-based reasoning.

\textbf{\techName{}-specific questionnaire.}
\techName{} received positive ratings across MS1-MS4, suggesting that participants considered its context interpretation and generated findings useful within their own analysis process.
MS2 (M=6.00) and MS4 (M=5.83) were rated highest, indicating that the surfaced findings were relevant to associated hypotheses and helped participants evaluate hypothesis support.
MS1 and MS3 received lower but still positive ratings (both M=5.67), suggesting that \techName{} usually captured the analysis context and produced useful findings, while participants still noticed variation in context interpretation and finding quality.
Together, these results support G2 by showing that \techName{} can surface relevant findings for hypothesis evaluation, while users still need to inspect and judge the generated findings.

\subsubsection{Quantitative Analysis}

\begin{table*}[t]
\footnotesize
\centering
\caption{Quantitative summary of generated hypotheses and findings. }
\label{tab:llm_output_quant}
\setlength{\tabcolsep}{1pt}
\begin{tabular}{@{}>{\raggedright\arraybackslash}m{0.14\textwidth}>{\raggedright\arraybackslash}m{0.16\textwidth}>{\centering\arraybackslash}m{0.05\textwidth}>{\raggedright\arraybackslash}m{0.418\textwidth}>{\centering\arraybackslash}m{0.142\textwidth}@{}}
\toprule
\textbf{Aspect} & \textbf{Measure} & \textbf{Total} & \textbf{Breakdown} & \textbf{Mean/Rate} \\
\midrule
\multirow{3}{*}{Output quantity} & Generated hypotheses & 26 & Origin: user-trace inferred/LLM-derived: 20/6 & 2.17/user \\
 & \multirow{2}{*}{Generated findings} & \multirow{2}{*}{293} & Source: user-trace inferred/LLM-generated: 181/112 & \multirow{2}{*}{24.42/user; 11.27/hyp.} \\
 & & & Type: visual/statistical/synthesized: 69/117/107 & \\
\midrule
Hypothesis alignment & All generated hypotheses & 26 & Yes/no/unsure: 25/0/1 & 96.2\% yes \\
Finding sufficiency & All generated hypotheses & 26 & Yes/partial/no/unsure: 17/7/1/1 & 92.3\% yes or partial \\
Finding relevance & All generated findings & 293 & Yes/no/unsure: 274/7/12 & 93.5\% yes \\
\bottomrule
\end{tabular}
\end{table*}
Table~\ref{tab:llm_output_quant} characterizes the LLM outputs from the 12 \techName{} sessions.
The framework produced 26 hypotheses (2.17/user) and 293 findings (24.42/user; 11.27/hypothesis), indicating that it translated participants' exploration into a substantial hypothesis-evidence structure rather than isolated LLM suggestions.
This source breakdown shows the intended co-analysis role: \techName{} not only organized participants' existing analysis into user-trace inferred hypotheses and findings (20 and 181), but also extended the current reasoning path through LLM-derived hypotheses and LLM-generated follow-up findings (6 and 112).
The finding breakdown further explains how the framework supported evidence-based reasoning.
The generated findings included visual (69), statistical (117), and synthesized findings (107), matching the design goal of combining inspectable view-based observations, computed evidence, and summaries across related findings.
The checklist results were positive, with high rates for hypothesis alignment (96.2\%), finding sufficiency or partial sufficiency (92.3\%), and finding relevance (93.5\%).
Overall, these results support G2 by showing that \techName{} organized and extended users' reasoning with relevant evidence for hypothesis evaluation.

\subsubsection{Qualitative Feedback}
Finally, we summarized the open-ended feedback from participants.
First, participants contrasted the analytical style of the two conditions: the baseline was lightweight, fast, and controllable, but also passive and linear, while \techName{} made the analysis ``more active and structured'' by suggesting follow-up directions and connecting hypotheses with evidence.
Second, participants valued evidence organization (U5, U9-U12); for example, U11 noted that ``hypotheses are linked to evidence,'' and others described how \techName{} connected holder links, temporal clusters, suspicious windows, repeated behaviors, and generated findings.
Third, participants also reported costs: \techName{} introduced more complexity and latency (U6, U8-U10), and some outputs included context not directly tied to the current hypothesis (U1, U9). U9 summarized this tradeoff as ``comprehensive analysis but sometimes contains some irrelevant context.''
Fourth, participants wanted more control and stronger critical support (U4-U7, U10-U12), including address search, drill-down into the aggregated \textit{Others} group, holder percentages, and clearer evidence attribution.
Several also expected the LLM to surface counterevidence, alternative explanations, and missing evidence; as U12 put it, the LLM should be ``more critical rather than only supportive.''

%% file: source/6_dicussion_conclusion.tex
\section{Discussion}

\subsection{Lessons for LLM-assisted Visual Analytics}
Our study suggests several lessons for designing LLM assistance in visual analytics.

\textbf{From reactive or proactive assistance to co-analysis.}
Recent LLM-assisted visual analytics systems mainly provide reactive or proactive support.
Reactive systems answer user requests or execute user-specified analysis steps~\cite{zhao2024lightva,ma2023insightpilot,zhao2024leva}, while proactive systems monitor interaction context and surface suggestions when help may be useful~\cite{zhao2025proactiveva}.
These systems still largely treat the user as the sole analyst and the LLM as an assistant to that process.
In contrast, \techName{} adopts the knowledge generation model for visual analytics~\cite{sacha2014knowledge} to position the LLM as a co-analyst within hypothesis-driven reasoning.
The LLM infers users' intent and existing findings from traces, identifies missing evidence, and performs follow-up visual and statistical analyses that support, refine, or contradict inferred hypotheses.
Participants' feedback also suggested that this design helped maintain users' domain judgment while reducing manual hypothesis-evidence iteration and strengthening analytical rigor.
How to combine reactive, proactive, and co-analysis support remains a promising direction for future LLM-assisted visual analytics systems.

\textbf{Supporting domain users requires flexibility in both visual analytics and LLM reasoning.}
Manipulation-risk assessment has no single criterion for correctness because manipulation strategies are diverse, evidence is partial, and interpretation depends on user assumptions~\cite{xiong2023market,eigelshoven2021cryptocurrency}.
Consistent with visual analytics as a knowledge-generation process~\cite{sacha2014knowledge}, users need to connect multiple evidence aspects, including holder concentration, relationship structure, suspicious behaviors, market movements, and alternative explanations.
Their risk preferences and expertise also affect how they configure heuristic parameters, how much detail they need for verification, and how they prefer to work with automated assistance.
This suggests that VA systems should support flexible evidence modeling and inspection, while LLM assistance should reason within the user's current context rather than assuming a fixed analytical standard.
In our user study, participants used the same system but followed different paths to findings; their ratings of hypothesis alignment and finding relevance suggest that \techName{} could, to some extent, adapt generated hypotheses and findings to these varied reasoning contexts.

\textbf{Balancing visual and statistical findings.}
Many existing LLM-assisted visual analytics systems generate insights through data-space analysis actions or use LLMs to produce visualization specifications and views~\cite{ma2023insightpilot,zhao2024lightva,dibia2023lida,qiu2025smartmlvs}.
In these systems, visualizations often serve as generated artifacts or user-facing summaries, while less attention has been paid to visualization-based LLM reasoning, where the LLM first generates views and then analyzes visual patterns for finding generation.
This raises a key design choice: whether the LLM should analyze data directly through scripts or first generate visualizations and inspect visual patterns.
Script-based analysis supports precise counting, aggregation, and parameter sweeping, and participants often described such statistical findings as making their analysis more solid and confident.
Visual findings are usually more qualitative: they stay close to the user's visual context and are easier to inspect, but depend on view quality and the LLM's visual interpretation ability.
\techName{} therefore combines both paths, using statistical actions for quantitative evidence and visual actions to connect findings back to inspectable views.
A useful future direction is to compare how LLMs perform in direct data analysis versus visualization-mediated pattern analysis, while considering how users understand, verify, and reuse the resulting findings.

\subsection{Generalizing the Collaborative Framework}
Although \techName{} is designed for cryptocurrency manipulation analysis, its framework may inform other hypothesis-driven visual analytics domains.
The knowledge generation model separates the computer-side components of visual analytics from the human-side reasoning process~\cite{sacha2014knowledge}.
Our framework follows this separation and makes two mappings explicit.
First, the \textit{data--model--visualization} mapping should be specified for the target domain.
In our system, the data are token transactions, the models include entity detection, relationship modeling, and suspicious-pattern flagging, and the visualizations include parameterized functions for the \textit{Token Distribution View}, \textit{Manipulation View}, and \textit{Behavior Detail View}.
Treating models and views as parameterized functions allows both users and LLMs to configure analysis settings, generate views, inspect results, and refer to the same evidence space.
Second, the \textit{action--finding--hypothesis} mapping should be specified for collaborative reasoning.
In our framework, user interactions and notes are captured as human actions, LLM operations are represented as analysis actions over model and visualization functions, and both are organized into findings that support, refine, or contradict hypotheses.
In this sense, generalization does not require transferring our domain rules directly.
Other domains can replace the data, models, views, and finding types while keeping the same separation between analytical substrate and reasoning structure.
A promising direction is to develop a general specification for LLM-assisted visual analytics, including criteria for defining domain models and view functions, deciding which actions an LLM can safely execute, and making generated findings auditable across domains.

\subsection{Limitations}
\techName{} is not without limitations.
First, the human-LLM collaborative framework can be slow because it involves hypothesis inference, analysis planning, and finding generation.
Executing suspicious-pattern flagging rules or generating visual findings can further increase response time.
To reduce the perceived waiting cost, \techName{} progressively displays intermediate results from each stage, and users can continue exploring the visual interface while the LLM analyzes in parallel.
Future systems could further reduce response time by allowing users to set analysis preferences, using faster LLM backends, and optimizing reasoning pipelines for common follow-up tasks.
Second, scalability remains challenging for large on-chain datasets with many transactions and holders.
Users may expect the LLM not only to support their current reasoning but also to search for potentially overlooked findings, which can increase both computational cost and visual complexity.
\techName{} mitigates this issue through filtering and on-demand detailed inspection, but views can still become overwhelming as the number of holders, relationships, and rule-flagged suspicious trading patterns grows.
More flexible filtering and preference controls could help users specify which holders, time ranges, or finding types should receive LLM attention and visual emphasis.

\section{Conclusion}
We presented \techName{}, an LLM-assisted visual analytics system for analyzing trade-based manipulation risks in cryptocurrency markets. \techName{} combines coordinated visualizations of holder distributions, holder relationships, suspicious trading patterns, detailed behaviors, and market dynamics with a human-LLM collaborative visual analytics framework for hypothesis inference, analysis planning, and finding generation. Two case studies and a user study showed that \techName{} supports manipulation-risk assessment, reduces repeated evidence seeking, and helps organize generated findings around users' hypotheses. In future work, we plan to further systematize the collaborative visual analytics framework and generalize it to broader application domains.

%% file: source/7_appendix.tex
\section{User Study Details}
\label{app:user_study_details}
\setcounter{table}{0}
\renewcommand{\thetable}{A\arabic{table}}
\renewcommand{\theHtable}{appendix.\arabic{table}}
\newcolumntype{Y}{>{\raggedright\arraybackslash}X}

\begin{table}[H]
\centering
\caption{Questionnaire, checklist, and interview items.}
\label{tab:appendix_questionnaire}
\scriptsize
\setlength{\tabcolsep}{4pt}
\renewcommand{\arraystretch}{0.98}
\begin{tabularx}{\textwidth}{@{}>{\raggedright\arraybackslash}p{0.07\textwidth}Y>{\raggedright\arraybackslash}p{0.12\textwidth}@{}}
\toprule
\textbf{Item} & \textbf{Question or checklist item} & \textbf{Format} \\
\midrule
\multicolumn{3}{@{}l}{\textit{Background questionnaire}} \\
\midrule
BQ1 & Age. & Numeric \\
BQ2 & Gender. & Single choice \\
BQ3 & Years of cryptocurrency trading or analysis experience. & Single choice \\
BQ4 & Background related to cryptocurrency, blockchain, or Web3. & Open text \\
BQ5 & Familiarity with on-chain transaction data. & 1-7 rating \\
BQ6 & Previous experience analyzing cryptocurrency manipulation or suspicious on-chain behaviors. & Multiple choice \\
BQ7 & Brief description of previous analysis experience, if any. & Open text \\
\midrule
\multicolumn{3}{@{}l}{\textit{Base visual analytics questionnaire}} \\
\midrule
VA1 & The Token Distribution View helped me understand the token holding distribution among major holders. & 1-7 Likert \\
VA2 & The Token Distribution View helped me inspect relationships among holders. & 1-7 Likert \\
VA3 & The Manipulation View helped me examine suspicious trading patterns in the market. & 1-7 Likert \\
VA4 & The Behavior Detail View helped me inspect detailed behaviors of selected holders. & 1-7 Likert \\
VA5 & Overall, the visual analytics system supported my manipulation-risk assessment. & 1-7 Likert \\
VA6 & The visual analytics system was easy to learn. & 1-7 Likert \\
VA7 & The visual analytics system was easy to use. & 1-7 Likert \\
\midrule
\multicolumn{3}{@{}l}{\textit{Post-condition LLM assistance questionnaire}} \\
\midrule
LLM1 & The LLM assistance reduced the amount of manual effort needed during the analysis. & 1-7 Likert \\
LLM2 & The LLM helped me find extra relevant evidence for my analysis goal. & 1-7 Likert \\
LLM3 & The LLM helped me organize evidence from different parts of the system. & 1-7 Likert \\
LLM4 & The LLM assistance increased my confidence that my analysis was supported by evidence. & 1-7 Likert \\
LLM5 & I trusted the LLM's responses during the analysis. & 1-7 Likert \\
\midrule
\multicolumn{3}{@{}l}{\textit{\techName{}-specific questionnaire}} \\
\midrule
MS1 & \techName{}'s interpretation of my analysis context accurately reflected what I was analyzing. & 1-7 Likert \\
MS2 & The new findings surfaced by \techName{} were relevant to their associated hypotheses. & 1-7 Likert \\
MS3 & I was satisfied with the quality of the generated findings. & 1-7 Likert \\
MS4 & Overall, the generated findings helped me evaluate whether the hypotheses were supported. & 1-7 Likert \\
\midrule
\multicolumn{3}{@{}l}{\textit{LLM output checklist}} \\
\midrule
OC1 & Whether each generated hypothesis was aligned with the participant's analysis context. & Yes/no/unsure \\
OC2 & Whether the associated findings were sufficient for evaluating each hypothesis. & Yes/part./no/unsure \\
OC3 & Whether each generated finding was relevant to its associated hypothesis. & Yes/no/unsure \\
\midrule
\multicolumn{3}{@{}l}{\textit{Open-ended feedback}} \\
\midrule
OE1 & If you gave any low score, please briefly explain the reason. & Open text \\
OE2 & How did you conduct the analysis task in the two conditions, and what were the pros and cons of \techName{} and the baseline? & Open text \\
OE3 & What suggestions do you have for improving \techName{}, including the visual analytics system and LLM assistance? & Open text \\
\bottomrule
\end{tabularx}
\end{table}

\begin{table}[H]
\centering
\caption{Participant background and counterbalanced assignment.}
\label{tab:appendix_counterbalance}
\scriptsize
\setlength{\tabcolsep}{2.5pt}
\renewcommand{\arraystretch}{1.12}
\resizebox{\textwidth}{!}{%
\begin{tabular}{@{}lcllcllll@{}}
\toprule
\textbf{User} & \textbf{Age} & \textbf{Gender} & \textbf{Exp.} & \textbf{Fam.} & \textbf{Background} & \textbf{Group} & \textbf{Session 1} & \textbf{Session 2} \\
\midrule
U1 & 26 & Male & 3-5 yrs & 6 & Finance; MEME trader; Web3 researcher & G1 & baseline/ACT & \techName{}/PNUT \\
U2 & 28 & Male & 1-2 yrs & 5 & Cryptocurrency investor & G1 & baseline/ACT & \techName{}/PNUT \\
U3 & 25 & Female & 3-5 yrs & 7 & Ph.D. blockchain transaction analysis & G1 & baseline/ACT & \techName{}/PNUT \\
U4 & 27 & Male & 3-5 yrs & 6 & Ph.D. smart-contract security & G2 & baseline/PNUT & \techName{}/ACT \\
U5 & 29 & Male & 3-5 yrs & 7 & Ph.D. research on cryptocurrency & G2 & baseline/PNUT & \techName{}/ACT \\
U6 & 31 & Female & $<$1 yr & 5 & Ph.D. research on blockchain & G2 & baseline/PNUT & \techName{}/ACT \\
U7 & 24 & Male & 1-2 yrs & 5 & RWA platform developer & G3 & \techName{}/ACT & baseline/PNUT \\
U8 & 27 & Male & 3-5 yrs & 5 & Cryptocurrency trader & G3 & \techName{}/ACT & baseline/PNUT \\
U9 & 25 & Female & 3-5 yrs & 7 & Ph.D. blockchain transaction analysis & G3 & \techName{}/ACT & baseline/PNUT \\
U10 & 23 & Male & 1-2 yrs & 5 & Web3 and cross-chain project developer & G4 & \techName{}/PNUT & baseline/ACT \\
U11 & 24 & Male & 1-2 yrs & 5 & Master's study in blockchain security; ChainGuard project & G4 & \techName{}/PNUT & baseline/ACT \\
U12 & 25 & Male & 1-2 yrs & 5 & MSc blockchain; Token2049 experience & G4 & \techName{}/PNUT & baseline/ACT \\
\bottomrule
\end{tabular}%
}
\vspace{2pt}
\begin{minipage}{0.98\textwidth}
\scriptsize
\textit{Note.} Exp. denotes cryptocurrency trading or analysis experience. Fam. denotes self-reported familiarity with on-chain transaction data on a seven-point scale.
\end{minipage}
\end{table}

\section{Prompts and Tools for Agents}
\label{app:prompts}

\subsection{Prompt Roles and Bounded Delegation}
\techName{} uses a layered prompt design to operationalize the human--LLM collaborative workflow described in Sec.~\ref{sec:llm_framework}. Rather than relying on a single monolithic assistant prompt, we separate interactive coordination, trace-level analysis, and evidence-specific follow-up into three agent roles:

\begin{itemize}
    \item \textit{Main chat agent.} This agent supports lightweight conversation with the user. It refreshes the current session context, distinguishes between direct questions, recommendation requests, full analysis, and incremental analysis, and keeps the interaction responsive while the user continues exploring the visual analytics interface.
    \item \textit{Analysis agent.} When a full or incremental analysis is necessary, the main chat agent delegates the long-running work to this background agent. The analysis agent owns the three-stage workflow in Fig.~\ref{fig:llmworkflow}: bottom-up hypothesis inference and generation, top-down analysis planning, and bottom-up finding generation.
    \item \textit{Task sub-agents.} The analysis agent may assign bounded sub-agents to independent branches of the analysis plan. Each task sub-agent receives a concrete investigation objective and returns evidence for the analysis agent to integrate.
\end{itemize}

Task sub-agents are defined by investigation objectives:
\begin{itemize}
    \item \textit{Support-seeking sub-agents} search for additional evidence for existing user hypotheses, analytic goals, or unresolved analytic questions. Depending on the evidence needed, the analysis agent is prompted to adapt one of three concrete sub-agent templates:
    \begin{itemize}
        \item \textit{Visual-analysis sub-agents} inspect recorded or newly rendered views for spatial, temporal, or behavioral patterns.
        \item \textit{Statistical-analysis sub-agents} compute exact counts, timestamps, amounts, overlaps, balances, or other quantitative evidence.
        \item \textit{Model-action robustness sub-agents} vary detector parameters, thresholds, or model outputs to test whether model-derived evidence remains stable.
    \end{itemize}
    If the checked evidence does not support the target, these sub-agents report the limitation rather than forcing a supporting finding.
    \item \textit{Skeptical sub-agents} search for counterevidence, benign alternatives, false positives, weak causal links, or qualifications that refine or contradict a hypothesis.
    \item \textit{Hypothesis-expansion sub-agents} test whether a derived or adjacent hypothesis has concrete support and should be supported, rejected, or deferred.
\end{itemize}

Each sub-agent is scoped to its assigned branch and cannot recursively delegate work. This bounded delegation keeps the chat agent available to the user while allowing the analysis agent to decompose complex hypothesis-evidence reasoning into parallel branches.

Table~\ref{tab:appendix_prompt_fragments} summarizes representative prompt fragments that instantiate these role distinctions.

\begin{table}[t]
\centering
\caption{Representative prompt fragments by agent role. The fragments are abbreviated to show the role logic rather than the full prompt text.}
\label{tab:appendix_prompt_fragments}
\scriptsize
\setlength{\tabcolsep}{3pt}
\renewcommand{\arraystretch}{1.08}
\begin{tabularx}{\textwidth}{@{}>{\raggedright\arraybackslash}p{0.20\textwidth}Y Y@{}}
\toprule
\textbf{Role} & \textbf{Role logic} & \textbf{Representative prompt fragment} \\
\midrule
Main chat agent & Coordinates the conversation, chooses the interaction mode, and keeps the interface responsive instead of performing long analyses inline. & ``You are embedded in the active ManiScope session. Work chat-first unless the user asks for durable outputs. Before trace-dependent answers, refresh the live trace and current state. Classify the request as lightweight chat, recommendation, full analysis, or incremental analysis. For heavy analysis, start a background analysis task, report its status, and let the user keep chatting.'' \\
\midrule
Analysis agent & Serves as the analysis owner. It connects the three-stage method into one coherent run and decides which branches should be delegated. & ``You are an independent background analysis agent, not the main chat agent. Complete the assigned analysis within the closed trace window. Reconstruct the user's reasoning, infer hypotheses, plan follow-up checks, generate visual, statistical, and synthesized findings, run skeptical review, and integrate or reject sub-agent outputs before reporting conclusions.'' \\
\midrule
Support-seeking Visual-analysis sub-agent & Gathers supporting visual findings when the claim depends on what can be inspected in ManiScope views. & ``You are a visual-analysis worker. Do not spawn additional agents. Analyze only the assigned trace window and target hypotheses. Use trace screenshots to reconstruct what the user saw, and render focused visual evidence when needed.'' \\
\midrule
Support-seeking Statistical-analysis sub-agent & Gathers supporting statistical findings when the claim depends on quantities that cannot be reliably read from the view alone. & ``You are a statistical-analysis worker. Do not spawn additional agents. Use raw data, model outputs, or scripts for exact counts, timestamps, amounts, overlaps, role statistics, profits or losses, and detector-output comparisons.'' \\
\midrule
Support-seeking Model-action robustness sub-agent & Checks whether support based on algorithmic outputs is robust to plausible model-configuration changes. & ``You are a model-action robustness worker. Do not spawn additional agents. Vary detector parameters, grouping rules, model outputs, thresholds, or linked components when relevant to test whether a model-derived claim is robust.'' \\
\midrule
Skeptical sub-agent & Challenges current hypotheses or high-level findings and returns qualifications, refinements, or contradictions when warranted. & ``You are a skeptical-review worker. Do not spawn additional agents. Search for negative evidence, false positives, benign alternatives, detector-parameter failures, missing causal links, denominator mistakes, or role-label ambiguities.'' \\
\midrule
Hypothesis-expansion sub-agent & Explores new hypothesis branches while preserving the distinction between evidence-backed hypotheses and speculative possibilities. & ``You are a hypothesis-expansion worker. Do not spawn additional agents. Test whether an adjacent hypothesis is supported by concrete follow-up findings rather than plausibility alone; if unsupported, rejected, or deferred, state why.'' \\
\bottomrule
\end{tabularx}
\end{table}

\subsection{Evidence Routing}
The prompts briefly define four evidence routes and ask agents to select the route before executing a planned task. These routes are orthogonal to the sub-agent objectives above: a support-seeking, skeptical, or hypothesis-expansion sub-agent may use visual, statistical, model-configuration, or synthesis actions depending on the claim being tested. Table~\ref{tab:appendix_evidence_routing} summarizes these routes with representative prompt fragments.

\begin{table}[t]
\centering
\caption{Representative prompt fragments for evidence routing.}
\label{tab:appendix_evidence_routing}
\scriptsize
\setlength{\tabcolsep}{3pt}
\renewcommand{\arraystretch}{1.08}
\begin{tabularx}{\textwidth}{@{}>{\raggedright\arraybackslash}p{0.18\textwidth}Y Y@{}}
\toprule
\textbf{Route} & \textbf{Purpose} & \textbf{Representative prompt fragment} \\
\midrule
Visual analysis & Produces visual findings from patterns that can be inspected in the interface or rendered views. & ``Use when a claim depends on spatial clusters, visible grouping, detector boundaries, links, card alignment, price-window alignment, behavior timelines, manipulation boxes, balance shapes, screenshots, rendered images, or values displayed by the GUI.'' \\
\midrule
Statistical analysis & Produces statistical findings from exact values, derived quantities, or computations beyond direct visual reading. & ``Use when a claim depends on exact counts, exact timestamps, exact amounts, transfer paths, wallet overlap, cohort market share, profit or loss, final balances, medians, means, detector-output overlap, or other derived values not displayed by the GUI.'' \\
\midrule
Model action & Tests claims that depend on detector outputs, grouping rules, or threshold-sensitive model results. & ``Use when the claim depends on detector outputs, suspicious labels, entity groups, manipulation boxes, link construction, component membership, or threshold-sensitive groupings.'' \\
\midrule
Synthesis action & Connects lower-level visual, statistical, or model evidence into synthesized findings. & ``Use when recording, comparing, qualifying, or connecting evidence already produced by visual, data, or model work.'' \\
\bottomrule
\end{tabularx}
\end{table}

\subsection{Trace Window Control}
Prompt design also controls the temporal scope of analysis. Because users may continue interacting with \techName{} while an agent is analyzing, an unconstrained agent could repeatedly absorb new actions, revise its plan, and make the evidence basis of a run ambiguous. We therefore scope each analysis run to the trace state available when the run starts. The prompt makes this boundary explicit:
\begin{quote}
\scriptsize
``Use this start anchor as the maximum trace boundary. Later live trace changes are out of scope and must be deferred to a later update analysis task.''
\end{quote}
This instruction turns trace-window control into an operational rule rather than a post-hoc explanation. Hypotheses and findings generated in one run are tied to the same interaction context, while later user actions become input for incremental analysis instead of silently changing the current run. This lets users continue exploring during long analyses while keeping the agent's reasoning auditable.

\subsection{Output Discipline}
Finally, the prompts impose output discipline so that agent results are inspectable rather than free-form summaries. Table~\ref{tab:appendix_output_discipline} summarizes the main types of output discipline and representative prompt fragments.

\begin{table}[t]
\centering
\caption{Representative prompt fragments for output discipline.}
\label{tab:appendix_output_discipline}
\scriptsize
\setlength{\tabcolsep}{3pt}
\renewcommand{\arraystretch}{1.08}
\begin{tabularx}{\textwidth}{@{}>{\raggedright\arraybackslash}p{0.20\textwidth}Y Y@{}}
\toprule
\textbf{Discipline} & \textbf{Purpose} & \textbf{Representative prompt fragment} \\
\midrule
Evidence grounding & Ensures that hypotheses and findings are tied to inspectable visual, statistical, model, or trace evidence. & ``Final conclusions must be grounded in trace evidence, rendered visual evidence, model output, data, or stated assumptions.'' \\
\midrule
Finding structure & Prevents isolated observations or redundant chains by organizing visual and statistical evidence into synthesized findings when needed. & ``Create a parent finding only when it adds synthesis, qualification, scope, contrast, uncertainty, or aggregation across evidence.'' \\
\midrule
Uncertainty and qualification & Encourages the agent to state limitations, caveats, and possible falsifying evidence rather than overclaiming support. & ``If a conclusion is uncertain, say what would confirm, weaken, or falsify it.'' \\
\midrule
Validation and integration & Requires the analysis agent to evaluate sub-agent outputs before adding them to the final reasoning. & ``Review sub-agent outputs. Reject weak branches, resolve conflicts, and integrate only verified candidate findings.'' \\
\midrule
User-facing presentation & Keeps final communication useful for analysts by separating evidence, open issues, and next-step implications. & ``Report completed checks, blocked checks, evidence, findings, and unresolved gaps.'' \\
\bottomrule
\end{tabularx}
\end{table}

Together, these rules ensure that generated outputs can be inspected, corrected, and reused by analysts during hypothesis verification.

\subsection{Tools for Agents}
Beyond prompts, \techName{} equips agents with tools that let them inspect the user context, execute analyses, and coordinate delegated work. We group these tools into seven categories:
\begin{itemize}
    \item \textit{Session and trace access} gives agents access to user interactions, annotations, screenshots, current visual state, and the trace window of an analysis run.
    \item \textit{Python execution} provides a general scripting environment for statistical analysis, raw-data inspection, and derived evidence computation.
    \item \textit{JavaScript/TypeScript execution} supports structured-output processing and validation utilities.
    \item \textit{Background analysis-agent control} lets the main chat agent start, inspect, and stop long-running analysis tasks without blocking the conversation.
    \item \textit{Sub-agent delegation} lets the analysis agent assign bounded evidence-gathering tasks to task sub-agents.
    \item \textit{Visualization and model-configuration tools} let agents render focused ManiScope views and vary relevant visual or detector configurations when evidence depends on them.
    \item \textit{Reasoning validation tools} let the analysis agent check consistency, integrate outputs, and reject weak or conflicting findings before presenting results.
\end{itemize}

Table~\ref{tab:appendix_agent_tool_access} summarizes how these tool categories are assigned across agent roles. ``Yes'' means the tool is accessible to the role, while ``Partial'' means access is limited by the assigned trace window, target hypothesis, or task branch.

\begin{table*}[t]
\centering
\caption{Tool access by agent role. Yes means accessible, not mandatory to use.}
\label{tab:appendix_agent_tool_access}
\scriptsize
\setlength{\tabcolsep}{4pt}
\renewcommand{\arraystretch}{1.08}
\resizebox{\textwidth}{!}{%
\begin{tabular}{@{}lccccccc@{}}
\toprule
\textbf{Agent role} & \textbf{Session/ trace} & \textbf{Python} & \textbf{JS/TS} & \textbf{Analysis control} & \textbf{Sub-agent delegation} & \textbf{Visualization/ model} & \textbf{Reasoning validation} \\
\midrule
Main chat agent & Yes & Yes & Yes & Yes & No & Yes & Yes \\
\midrule
Analysis agent & Yes & Yes & Yes & No & Yes & Yes & Yes \\
\midrule
Support-seeking visual-analysis sub-agent & Partial & Yes & Yes & No & No & Yes & Partial \\
\midrule
Support-seeking statistical-analysis sub-agent & Partial & Yes & Yes & No & No & Partial & Partial \\
\midrule
Support-seeking model-action robustness sub-agent & Partial & Yes & Yes & No & No & Yes & Partial \\
\midrule
Skeptical sub-agent & Partial & Yes & Yes & No & No & Partial & Partial \\
\midrule
Hypothesis-expansion sub-agent & Partial & Yes & Yes & No & No & Partial & Partial \\
\bottomrule
\end{tabular}%
}
\end{table*}